\documentclass[prd,nofootinbib,preprint,superscriptaddress]{revtex4}

\usepackage{amsmath}
\usepackage[dvips]{graphicx}

\newcommand{\capdef}{}
\newcommand{\mycaption}[2][\capdef]{\renewcommand{\capdef}{#2}%
       \caption[#1]{{\footnotesize #2}}}
\makeatletter
\renewcommand{\fnum@table}{\textbf{\tablename~\thetable}}
\renewcommand{\fnum@figure}{\textbf{\figurename~\thefigure}}
\makeatother

\newcommand{\Dmq}{\Delta m^2}
\newcommand{\eVq}{\ensuremath{\text{eV}^2}}
\newcommand{\diag}{\mathop{\mathrm{diag}}}

\begin{document}


\vspace*{10mm}

\title{Sterile neutrino oscillations after first MiniBooNE results}

\author{Michele Maltoni}
\email{maltoni@delta.ft.uam.es}
\affiliation{Departamento de F\'isica Te\'orica \& Instituto de
  F\'isica Te\'orica, Facultad de Ciencias C-XI, Universidad
  Aut\'onoma de Madrid, Cantoblanco, E-28049 Madrid, Spain}

\author{Thomas Schwetz} 
\email{schwetz@cern.ch}
\affiliation{Physics Department, Theory Division, CERN, CH--1211
  Geneva 23, Switzerland}

\vspace*{.5cm}

\begin{abstract}
\vspace*{.5cm} In view of the recent results from the MiniBooNE experiment we
    revisit the global neutrino oscillation fit to short-baseline neutrino
    data by adding one or two sterile neutrinos with eV-scale masses to the
    three Standard Model neutrinos, and for the first time we consider also
    the global fit with three sterile neutrinos. Four-neutrino oscillations of
    the (3+1) type have been only marginally allowed before the recent
    MiniBooNE results, and become even more disfavored with the new data (at
    the level of $4\sigma$). 
    In the framework of so-called (3+2) five-neutrino mass schemes we find
    severe tension between appearance and disappearance experiments at the
    level of more than $3\sigma$, and hence no satistfactory fit to the global
    data is possible in (3+2) schemes. This tension remains also when a third
    sterile neutrino is added, and the quality of the global fit does not
    improve significantly in a (3+3) scheme. It should be noted, however, that
    in models with more than one sterile neutrino the MiniBooNE results are in
    perfect agreement with the LSND appearance evidence, thanks to the
    possibility of CP violation available in such oscillation
    schemes. Furthermore, if disappearance data are not taken into account
    (3+2) oscillations provide an excellent fit to the full MiniBooNE spectrum
    including the event excess at low energies.
\end{abstract}

\preprint{CERN-PH-TH/2007-075}
\preprint{IFT-UAM/CSIC-07-19}

\maketitle


\section{Introduction}

Recently first results from the MiniBooNE (MB) experiment~\cite{MB,
MB-talk} at Fermilab have been released on a search for
$\nu_\mu\to\nu_e$ appearance with a baseline of 540~m and a mean
neutrino energy of about 700~MeV. The primary purpose of this
experiment is to test the evidence of $\bar\nu_\mu \to \bar\nu_e$
transitions reported by the LSND experiment at Los
Alamos~\cite{Aguilar:2001ty} with a very similar $L/E$ range.
Reconciling the LSND signal with the other evidence for neutrino
oscillations is a long-standing challenge for neutrino phenomenology,
since it requires a mass-squared difference at the \eVq\ scale, at odd
with the values needed to explain atmospheric~\cite{sk-atm},
long-baseline accelerator~\cite{Aliu:2004sq, Michael:2006rx},
solar~\cite{sol-radiochem, SK-solar, sno}, and long-baseline
reactor~\cite{kamland} neutrino data.

It turns out that introducing a fourth (sterile)
neutrino~\cite{sterile} does not lead to a satisfactory description of
all data in terms of neutrino oscillations~\cite{Maltoni:2002xd,
strumia} because of tight constraints from atmospheric~\cite{sk-atm},
solar~\cite{sno}, and null-result short-baseline (SBL)
experiments~\cite{karmen, Astier:2003gs, Dydak:1983zq, Declais:1994su}
(see Ref.~\cite{early} for early four-neutrino analyses
considering LSND, and Refs.~\cite{Maltoni:2004ei,
Gonzalez-Garcia:2007ib} for recent updates). So-called (2+2) schemes
are ruled out by strong constraints on a sterile neutrino component in
solar as well as in atmospheric neutrino
oscillations~\cite{sol-atm-4nu} at high significance. Therefore, we
will not consider such schemes in the following. Also (3+1) schemes
suffer from a well-known tension between the LSND appearance signal
and null-result SBL disappearance experiments~\cite{Bilenky:1996rw,
Okada:1996kw, Barger:1998bn, Bilenky:1999ny, Peres:2000ic,
Grimus:2001mn, cornering}. We will show that recent MB results further
aggravate this tension and hence (3+1) schemes get even more
disfavoured.
In Ref.~\cite{Sorel:2003hf} a five-neutrino mass scheme of the (3+2)
type has been considered, arguing that the disagreement between LSND
and null-result experiments becomes somewhat relaxed compared to
(3+1), see also Ref.~\cite{Peres:2000ic}. Here we reconsider this
possibility in view of the recent MB data. Furthermore, we investigate
the impact of adding a third sterile neutrino on the quality of the
global fit. Since we know that there are three active neutrinos, the
possibility of three sterile neutrinos is appealing for aesthetical
reasons.

Apart from sterile neutrino oscillations, various more exotic
explanations of the LSND signal have been proposed, for example,
neutrino decay~\cite{Ma:1999im, Palomares-Ruiz:2005vf}, CPT
violation~\cite{cpt, strumia}, violation of Lorentz
symmetry~\cite{lorentz}, a lepton number violating muon
decay~\cite{LNV}, CPT-violating quantum
decoherence~\cite{Barenboim:2004wu}, mass-varying
neutrinos~\cite{MaVaN}, or shortcuts of sterile neutrinos in extra
dimensions~\cite{Pas:2005rb}.

In this work we concentrate on the oscillation framework including one
or more sterile neutrinos at the eV scale. Such models do have an
impact on cosmology~\cite{Cirelli:2004cz}. First, the sterile
neutrinos will contribute to the effective number of neutrino species
at Big Bang nucleosynthesis~\cite{sterileBBN, Okada:1996kw}, and
second, these models are subject to strong bounds on the sum of the
neutrino masses in the sub-eV range from the combination of various
cosmological data sets (see \textit{e.g.}, Ref.~\cite{cosmo-bounds}
for recent analyses). In order to reconcile such neutrino schemes with
cosmology some non-standard scenario has to be invoked, see for
example Refs.~\cite{L-asymm, BBN-majoron, Gelmini:2004ah}.

The outline of the paper is as follows. In Sec.~\ref{sec:3+1} we
consider (3+1) four-neutrino schemes, and give some details on the
used data and their analysis. Sec.~\ref{sec:3+2} is devoted to (3+2)
five-neutrino schemes, discussing the compatibility of LSND and MB in
such schemes, as well as the problems of these models to reconcile
appearance and disappearance experiments. In Sec.~\ref{sec:3+3} we
extend the (3+2) scheme by adding a third sterile neutrino to a (3+3)
six-neutrino model and investigate whether the global fit improves
significantly. We summarise in Sec.~\ref{sec:summary}. In
appendix~\ref{app:CP} we discuss the mechanism to reconcile LSND and
MB by CP violation in (3+2) schemes, appendix~\ref{app:atm} contains a
parameter counting of general sterile neutrino oscillation schemes,
and in appendix~\ref{app:robust} we consider details of the analysis
of atmospheric neutrino data in such models.


\section{(3+1) four-neutrino mass schemes}
\label{sec:3+1}

The (3+1) four-neutrino spectra are a small perturbation to the
standard three-active neutrino case. A cluster of three neutrino mass
states accounts for the ``solar'' ($\Dmq_{21}$) and ``atmospheric''
($\Dmq_{31}$) mass splittings. The fourth mass state is separated by
an eV-scale mass gap to account for the LSND oscillations, and there
is only small mixing of active neutrinos with this mass eigenstate.


\subsection{Appearance data in (3+1) schemes}

We start our discussion of the (3+1) mass scheme by considering the
SBL appearance experiments in the $\nu_\mu\to\nu_e$ (or
$\bar\nu_\mu\to\bar\nu_e$) channel, including the
LSND~\cite{Aguilar:2001ty} evidence, the bounds from
KARMEN~\cite{karmen} and NOMAD~\cite{Astier:2003gs}, and the recent
MB~\cite{MB} data. A combined analysis of LSND and KARMEN can be found
in Ref.~\cite{Church:2002tc}.
In the approximation $\Dmq_{21} \approx \Dmq_{31} \approx 0$, the SBL
appearance probability in (3+1) schemes is equivalent to the
two-neutrino case, where the effective mixing angle is determined by
$\sin^2 2\theta_\text{SBL} = 4|U_{e4}|^2 |U_{\mu 4}|^2$. Therefore, the
analysis performed by the MB collaboration~\cite{MB, MB-talk} directly
applies to (3+1) schemes. We comment only briefly on this case, with
the main purpose to check our analysis against the official MB
results.

For our re-analysis of LSND we fit the observed transition
probability (total rate) plus 11 data points of the $L/E$ spectrum
with free normalisation, both derived from the decay-at-rest
data~\cite{Aguilar:2001ty}. For KARMEN the data observed in 9 bins of
prompt energy as well as the expected background~\cite{karmen} is used
in the fit.  Further details of our LSND and KARMEN analyses are given
in Ref.~\cite{Palomares-Ruiz:2005vf}. For NOMAD we fit the total rate
using the information provided in Ref.~\cite{Astier:2003gs}; our
exclusion curve is in good agreement with the result presented in that
reference.

The MB re-analysis is based on the $\nu_\mu$ neutrino flux,
efficiencies, and energy resolution provided in Ref.~\cite{MB-talk},
folded with the $\nu_e$ charged-current quasi-elastic (CCQE) cross
section, to obtain a prediction for the CCQE event excess from
$\nu_\mu\to\nu_e$ oscillations. We calibrate our simulation to the
official MB analysis using the prediction for two example points
provided in Ref.~\cite{MB}.
For the fit the spectrum of excess events binned in reconstructed
neutrino energy from Fig.~2 of Ref.~\cite{MB} is used, where the error
bars include statistical errors and the uncertainty from the
background prediction. 
Detailed technical information the MB oscillation analysis is
available at the webpage Ref.~\cite{MB-data}, including efficiencies
and error correlations. Our MB results are in good agreement with the
official MB analysis as described in that webpage.

MB data are consistent with zero (no excess) above 475~MeV, whereas
below this energy a $3.6\sigma$ excess of $96 \pm 17 \pm 20$ events is
observed. Whether this excess comes indeed from $\nu_\mu\to\nu_e$
transitions or has some other origin is under
investigation~\cite{MB}. Lacking any explanation in terms of
backgrounds or systematical uncertainties we take these data at face
value, and in some cases we will use all 10 bins of the full energy
range from 300~MeV to 3~GeV in the fit (``MB300'').
However, as discussed in Refs.~\cite{MB, MB-talk}, two-neutrino
oscillations cannot account for the event excess at low energies. We
confirm that the quality of the (3+1) MB fit drastically worsens when
the two energy bins between 300 and 475~MeV are included in the fit.
Therefore, we follow the strategy of the MB collaboration and restrict
the (3+1) analysis to the energy range from 475~MeV to 3~GeV
(``MB475'').

\begin{figure}[t] \centering 
    \includegraphics[width=0.54\textwidth]{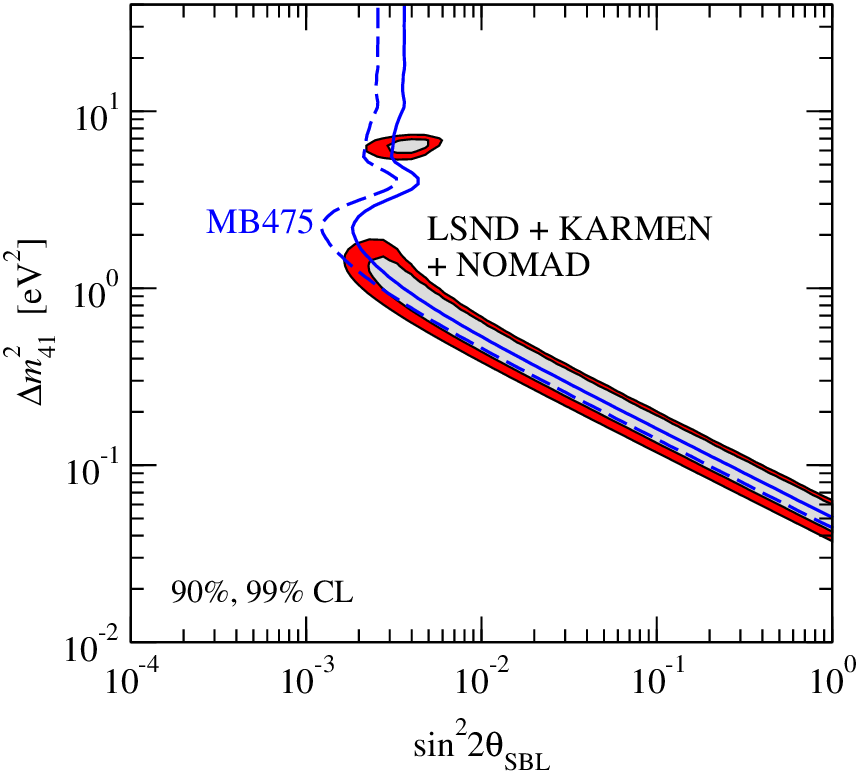}
    \mycaption{\label{fig:4nu-app}%
      Allowed region for MB475 (solid and dashed curves) and
      LSND+KARMEN+NOMAD (shaded regions) at 90\% and 99\%~CL (2 dof)
      in (3+1) mass schemes.}
\end{figure}

The bound from MB475 data is shown in Fig.~\ref{fig:4nu-app} in
comparison with the allowed region from the combined LSND, KARMEN,
NOMAD data. In agreement with Refs.~\cite{MB, MB-talk} we find that
the 90\%~CL regions do not overlap. A marginal overlap appears if both
data sets are stretched to the 99\%~CL. If all data are summed we find
a best fit point with $\chi^2_\text{min} = 26.6$ for $(29-2)$ degrees
of freedom (dof). Although this leads to a very good nominal
goodness-of-fit (gof), the figure clearly shows that there is
significant tension between MB and LSND.
A powerful tool to evaluate the compatibility of different data sets
is the so-called parameter goodness-of-fit (PG) criterion discussed in
Ref.~\cite{Maltoni:2003cu}. It is based on the $\chi^2$ function
\begin{equation} \label{eq:PG}
    \chi^2_\text{PG} = 
    \chi^2_\text{tot,min} - \sum_i \chi^2_{i,\text{min}} \,,
\end{equation}
where $\chi^2_\text{tot,min}$ is the $\chi^2$ minimum of all data sets
combined and $\chi^2_{i,\text{min}}$ is the minimum of the data set
$i$. This $\chi^2$ function measures the ``price'' one has to pay by
the combination of the data sets compared to fitting them
independently. It should be evaluated for the number of dof
corresponding to the number of parameters in common to the data sets,
see Ref.~\cite{Maltoni:2003cu} for a precise definition.
Applying this test to check the compatibility of MB with the other SBL
appearance data, we find $\chi^2_\text{PG} = 7.4$ (2~dof),
corresponding to a PG of 2.5\%. If we test the compatibility of the
three data sets MB, LSND, and KARMEN+NOMAD we find $\chi^2_\text{PG} =
13.7$ (4~dof), and $\text{PG} = 0.8\%$. In the latter case the slight
tension between LSND and KARMEN also contributes to the $\chi^2$,
whereas in the first case this tension is removed since they are added
into one single data set.


\subsection{Global SBL data in (3+1) schemes}
\label{sec:global-31}

Now we proceed to the global four-neutrino analysis, adding also the
information from disappearance experiments. We include the
Bugey~\cite{Declais:1994su}, Chooz~\cite{Apollonio:2002gd}, and Palo
Verde~\cite{Boehm:2001ik} $\bar\nu_e$ reactor experiments, as well as
the CDHS~\cite{Dydak:1983zq} $\nu_\mu$ disappearance experiment.
Details of our Bugey and CDHS fits can be found in
Ref.~\cite{Grimus:2001mn}. Furthermore, atmospheric neutrino data give
an important constraint on the parameter $d_\mu$~\cite{Bilenky:1999ny,
cornering}, which is equal to $|U_{\mu 4}|^2$ in (3+1) schemes, see
Ref.~\cite{MSV-4nu} and appendices~\ref{app:atm},
\ref{app:robust}. This information is crucial for the region of
$\Dmq_{41} \lesssim 1~\eVq$, where the bound from CDHS disappears. We
use the updated atmospheric neutrino analysis from
Ref.~\cite{Gonzalez-Garcia:2007ib}, which includes also recent
K2K~\cite{Aliu:2004sq} and MINOS~\cite{Michael:2006rx} data, and
include the bound on $d_\mu$ as one single data point in the SBL fit.
The total number of data points in the (3+1) analysis is
\begin{equation} \label{eq:data}
    \left.\begin{aligned}
	N_\text{APP} &=
	11_\text{(LSND)} + 9_\text{(KARMEN)} + 1_\text{(NOMAD)} +
	8_\text{(MB475)} = 29
	\\
	N_\text{DIS} &=
	60_\text{(Bugey)} + 1_\text{(Chooz)} + 1_\text{(Palo Verde)} +
	15_\text{(CDHS)} + 1_\text{(ATM)} = 78
    \end{aligned}\right\rbrace \quad
    N_\text{tot} = 107 \,.
\end{equation}

It is well known that (3+1) schemes suffer from a tension between the
LSND appearance signal and the bounds from disappearance experiments,
see, \textit{e.g.}, Refs.~\cite{Bilenky:1996rw, Okada:1996kw,
Barger:1998bn, Bilenky:1999ny, Peres:2000ic, Grimus:2001mn}. Reactor
experiments constrain the parameter $|U_{e4}|^2 \ll 1$, CDHS and
atmospheric neutrinos limit $|U_{\mu4}|^2 \ll 1$, whereas the LSND
oscillation amplitude is given by $4 |U_{e4}|^2|U_{\mu4}|^2$. Because
of this tension (3+1) schemes have been already disfavoured before the
recent MB results, see Ref.~\cite{Maltoni:2004ei}. Testing the
compatibility of LSND with all the no-evidence experiments (NEV)
without MB leads to a $\chi^2_\text{PG} = 20.9$ for 2~dof, which
indicates an inconsistency at high CL. Since MB475 data is also in
conflict with the LSND signal the new data adds to the tension and we
find
\begin{equation}\label{eq:PG4nu}
    \chi^2_\text{PG} = 24.7 ~\text{(2 dof)} \,,
    \qquad \text{LSND vs NEV (incl. MB475)} \,.
\end{equation}
Alternatively one may test the compatibility of the three data sets
LSND, MB475+KARMEN+NOMAD (NEV-APP), and disappearance experiments,
which gives
\begin{equation}
    \chi^2_\text{PG} = 24.8 ~\text{(4 dof)} \,,
    \qquad \text{LSND vs NEV-APP vs DIS} \,.
\end{equation}
These numbers formally imply inconsistency at more than $4\sigma$, and
hence we conclude that (3+1) schemes are highly disfavoured by recent
data.  This level of incompatibility is already close to the one of
solar and atmospheric neutrino data in (2+2) four-neutrino
schemes~\cite{Maltoni:2002xd, Maltoni:2004ei, Gonzalez-Garcia:2007ib}
with a $\chi^2_\text{PG,(2+2)} = 30.7$ (1~dof). Let us mention that
the global best fit point of all data has $\chi^2_\text{min} =
100.7$ for $(107-3)$~dof, which nominally gives an excellent
gof. However, in the standard gof test the problems in the fit get
diluted by the large number of data points, and the disagreement of
different data sets becomes only visible when they are analysed
separately and then compared to each other, see
Ref.~\cite{Maltoni:2003cu} for a detailed discussion.

\begin{figure}[t] \centering
    \includegraphics[width=0.54\textwidth]{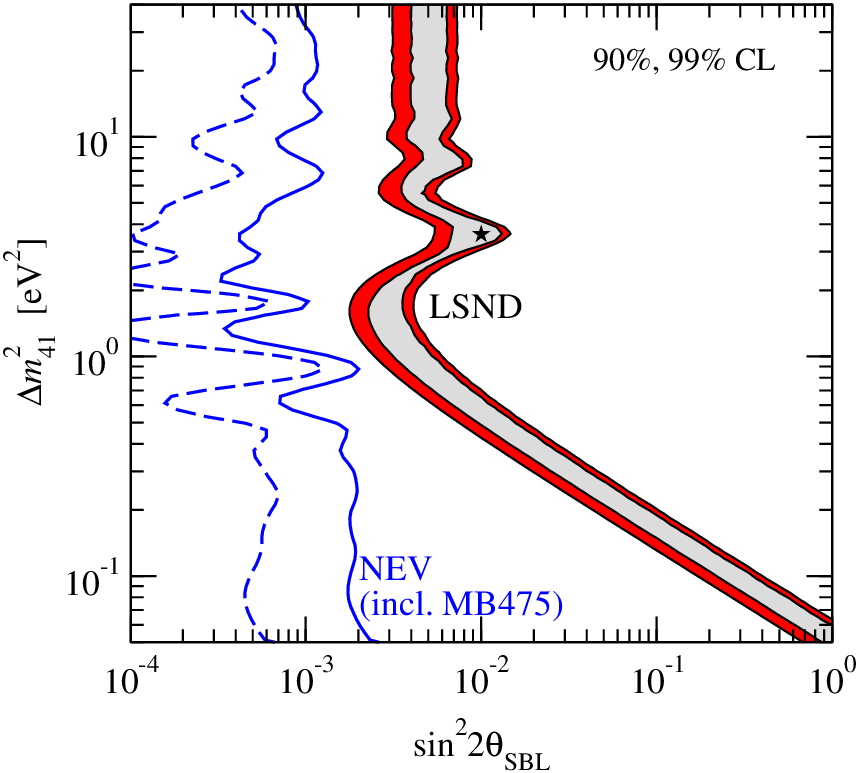}
    \mycaption{\label{fig:4nu-bound}%
      Allowed regions in (3+1) schemes from no-evidence (NEV) data
      including MB475 (solid and dashed curves) and LSND (shaded
      regions) at 90\% and 99\%~CL (2~dof).}
\end{figure}

The bound from NEV data in comparison with the LSND region is shown in
Fig.~\ref{fig:4nu-bound}. Let us mention that the shape of the NEV
bound does hardly change by adding MB data, however the $\chi^2$ of
the global best fit point increases significantly (see above). Note
also, that for our LSND analysis we use only the decay-at-rest data,
where the appearance signal is most significant.\footnote{The LSND
allowed region of Fig.~\ref{fig:4nu-bound} consists of a connected
band, which shows that the fit is dominated by the total rate and the
spectral information available to us is not strong enough to produce
disconnected regions as obtained from an event-based likelihood
analysis~\cite{Church:2002tc}. However, the location and size of our
region are in very good agreement with Ref.~\cite{Church:2002tc}, and
moreover, the regions of the parameter space which are relevant for
the combination with NEV data are reproduced very well.} If the global
LSND data including also the decay-in-flight events are used, the LSND
regions shift to slightly smaller values of $\sin^22\theta_\text{SBL}$
and the disagreement with NEV gets somewhat less severe, see
Ref.~\cite{Maltoni:2002xd} for a discussion of this issue.

To summarise, (3+1) schemes get further disfavoured by MB data for two
reasons. First, in these models LSND and MB are in disagreement at the
level of 90\% to 98.5\%~CL~\cite{MB, MB-talk}, and second, the tension
between LSND and NEV data gets worse due to MB.


\section{(3+2) five-neutrino mass schemes}
\label{sec:3+2}

Five-neutrino schemes of the (3+2) type are a straight-forward
extension of (3+1) schemes. In addition to the cluster of the three
neutrino mass states accounting for ``solar'' and ``atmospheric'' mass
splittings now two states at the eV scale are added, with a small
admixture of $\nu_e$ and $\nu_\mu$ to account for the LSND signal.  In
Ref.~\cite{Sorel:2003hf} it has been argued that in (3+2) schemes the
tension between LSND and NEV data becomes significantly relaxed
compared to the (3+1) case. Here we reconsider this possibility in the
light of the new MB data.


\subsection{Appearance data in (3+2) schemes}
\label{sec:app-32}

First we consider appearance data only (LSND, KARMEN, NOMAD, and MB).
In the SBL approximation $\Dmq_{21} \approx \Dmq_{31} \approx 0$, the
relevant appearance probability is given by
\begin{multline} \label{eq:5nu-prob}
    P_{\nu_\mu\to\nu_e} =
    4 \, |U_{e4}|^2 |U_{\mu 4}|^2 \, \sin^2 \phi_{41} +
    4 \, |U_{e5}|^2 |U_{\mu 5}|^2 \, \sin^2 \phi_{51}
    \\
    + 8 \,|U_{e4}U_{\mu 4}U_{e5}U_{\mu 5}| \,
    \sin\phi_{41}\sin\phi_{51}\cos(\phi_{54} - \delta) \,,
\end{multline}
with the definitions
\begin{equation} \label{eq:5nu-def}
    \phi_{ij} \equiv \frac{\Dmq_{ij}L}{4E} \,,
    \qquad \delta \equiv
    \arg\left(U_{e4}^* U_{\mu 4} U_{e5} U_{\mu 5}^* \right) \,. 
\end{equation}
Eq.~\eqref{eq:5nu-prob} holds for neutrinos (NOMAD and MB); for
anti-neutrinos (LSND and KARMEN) one has to replace $\delta \to
-\delta$. Since Eq.~\eqref{eq:5nu-prob} is invariant under the
transformation $4\leftrightarrow 5$ and $\delta \to -\delta$, we can
restrict the parameter range to $\Dmq_{54} \ge 0$, or equivalently
$\Dmq_{51} \ge \Dmq_{41}$, without loss of generality. Note also that
the probability Eq.~\eqref{eq:5nu-prob} depends only on the
combinations $|U_{e4}U_{\mu 4}|$ and $|U_{e5}U_{\mu 5}|$, and
therefore, the total number of independent parameters is 5 if only
appearance experiments are considered.

\begin{figure}[t] \centering 
    \includegraphics[width=0.9\textwidth]{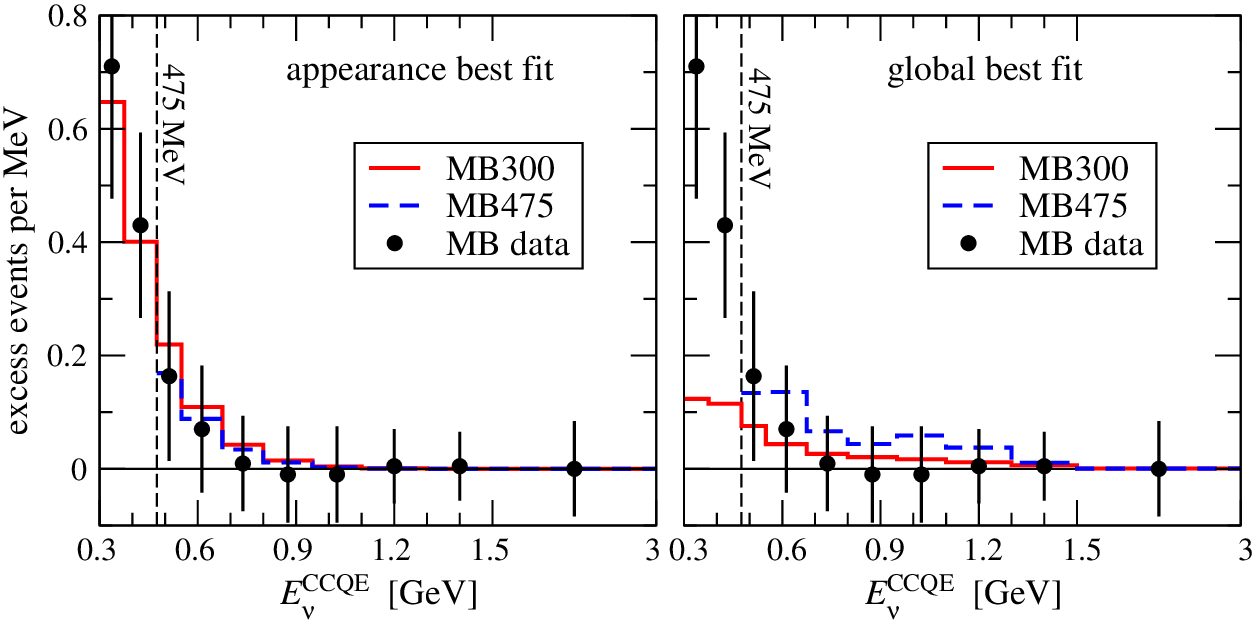}
    \mycaption{\label{fig:spectrum}%
      MB spectral data in bins of reconstructed CCQE neutrino energy.
      The histograms show the prediction at the best fit points in
      (3+2) mass schemes for SBL appearance data LSND, KARMEN, NOMAD,
      MB (left), and for the global data (right). For the solid
      histograms the full MB energy range has been used in the fit
      (MB300), whereas for the dashed histogram the two lowest energy
      data points have been omitted (MB475). The corresponding
      parameter values are given in Tab.~\ref{tab:bfp}.}
\end{figure}

Non-trivial values of the complex phase $\delta$ lead to CP violation,
and hence in (3+2) schemes much more flexibility is available to
accommodate the results of LSND (anti-neutrinos) and MB
(neutrinos).\footnote{The possibility to use CP violation to reconcile
LSND with a possible null-result of MB neutrino data was pointed out
in Ref.~\cite{Palomares-Ruiz:2005vf} in the framework of neutrino
decay, and later in Ref.~\cite{Karagiorgi:2006jf} in relation with
(3+2) oscillations.}
Indeed we find that MB is perfectly compatible with LSND in the (3+2)
framework. In Fig.~\ref{fig:spectrum} (left) we show the prediction
for MB at the best fit points in the combined MB, LSND, KARMEN, NOMAD
analysis.  Clearly MB data can be fitted very well by simultaneously
explaining the LSND evidence; we have checked that the prediction for
the LSND oscillation probability is within the $1\sigma$ range of the
observed value. In this case also the low energy MB data can be
explained, and therefore, in contrast to (3+1) schemes, (3+2)
oscillations offer an appealing possibility to account for this
excess.  In the following we will present results from both MB data
sets, MB475 as well as MB300.
Note that for MB475 the number of data points used in our analysis is
given in Eq.~\eqref{eq:data}, whereas for the case of MB300 two more
data points should be added to $N_\text{APP}$ and $N_\text{tot}$.
The parameter values and the $\chi^2$ minima at the best fit points
are given in Tab.~\ref{tab:bfp}. In both cases, MB475 and MB300, a gof
of 85\% is obtained, showing that MB is in very good agreement with
global SBL appearance data including LSND. 

\begin{table}[t] \centering
    \begin{tabular}{l@{\quad}ccc@{\quad}ccc@{\quad}c@{\quad}c@{\quad}c}
	\hline\hline
	data set
	& \multicolumn{2}{c}{$|U_{e4} U_{\mu 4}|$} & $\Dmq_{41}$ 
	& \multicolumn{2}{c}{$|U_{e5} U_{\mu 5}|$} & $\Dmq_{51}$  
	& $\delta$ & $\chi^2_\text{min} / \text{dof}$ & gof
	\\
	\hline
	appearance (MB475) &
	\multicolumn{2}{c}{0.044} & 0.66 &
	\multicolumn{2}{c}{0.022} & 1.44 & 
	1.12$\pi$ & $16.9/(29 - 5)$ & 85\%
	\\
	appearance (MB300) &
	\multicolumn{2}{c}{0.31} & 0.66 &
	\multicolumn{2}{c}{0.27} & 0.76 & 
	1.01$\pi$ & $18.5/(31 - 5)$  & 85\%
	\\
	\hline
	& $|U_{e4}|$ & $|U_{\mu 4}|$ & 
	& $|U_{e5}|$ & $|U_{\mu 5}|$ & 
	& &
	\\
	\hline
	global data (MB475) & 
	0.11 & 0.16 & 0.89 &
	0.12 & 0.12 & 6.49 &
	1.64$\pi$ & $94.5/(107 - 7)$ & 63\%
	\\
	global data (MB300) & 
	0.12 & 0.18 & 0.87 &
	0.11 & 0.089 & 1.91 &
	1.44$\pi$ & $104.4/(109 - 7)$ & 41\%
	\\
	\hline\hline
    \end{tabular}
    \mycaption{\label{tab:bfp}%
      Parameter values, $\chi^2$, and gof of the best fit points for
      SBL appearance data from LSND, KARMEN, NOMAD, MB (upper part),
      and global data (lower part) in (3+2) schemes. Mass-squared
      differences are given in eV$^2$. Results are shown without
      (MB475), and including (MB300) the low energy data from MB.}
\end{table}

\begin{figure}[t] \centering
    \includegraphics[width=0.6\textwidth]{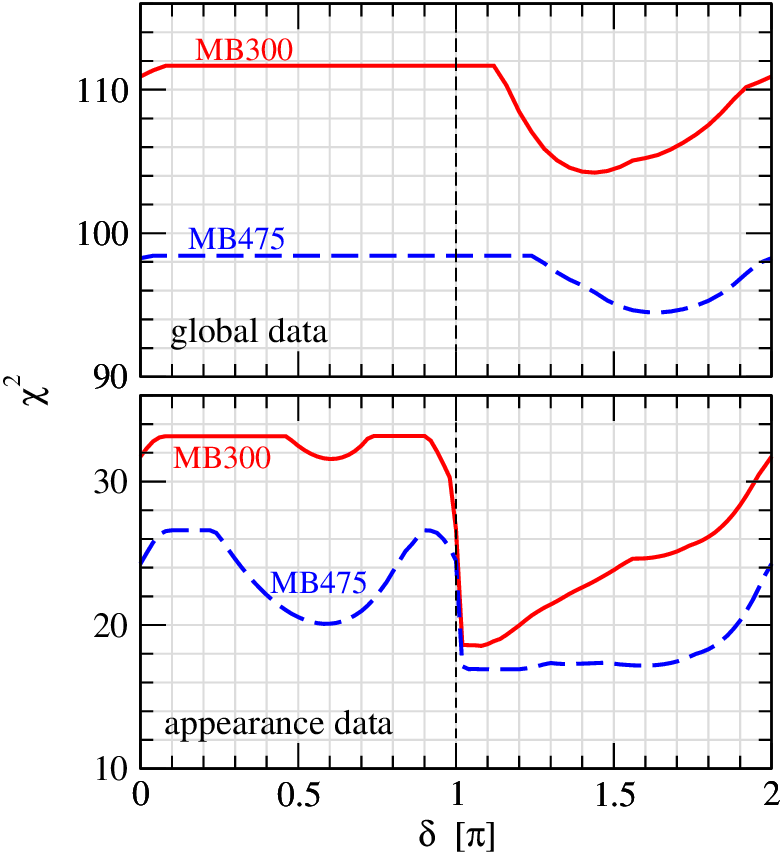}
      \mycaption{\label{fig:chisq_delta}%
	The $\chi^2$ in (3+2) schemes as a function of the CP phase
	$\delta$ defined in Eq.~\eqref{eq:5nu-def} for appearance data
	from LSND, KARMEN, NOMAD, and MB (bottom), and for global data
	(top). Results are shown without (MB475), and including
	(MB300) the low energy data from MB. All other parameters have
	been minimised, respecting the constraint $\Dmq_{51} \ge
	\Dmq_{41}$.}
\end{figure}

In Fig.~\ref{fig:chisq_delta} (bottom) the $\chi^2$ is shown as a
function of the CP phase $\delta$. The data prefer values in the range
$\pi < \delta < 2\pi$ in order to reconcile LSND and MB. However, as
visible in the figure no pronounced minimum appears and a rather broad
range of $\delta$ values lead to a good fit, including also values
rather close to the CP conserving value $\delta = \pi$. For MB300 the
best fit even occurs at $\delta = 1.01\pi$, compare
Tab.~\ref{tab:bfp}. In appendix~\ref{app:CP} we give an explanation
how LSND and MB can be reconciled with values of $\delta$ that close
to $\pi$. As discussed there, this mechanism is based on a delicate
interplay of the three terms in the probability of
Eq.~\eqref{eq:5nu-prob}, and a ``tiny amount of CP violation'' suffice
to make LSND and MB compatible. However, this mechanism requires
rather large values of the appearance amplitudes $|U_{e4} U_{\mu 4}|$ and
$|U_{e5} U_{\mu 5}|$, which are not compatible with disappearance
data, and hence solutions with $\delta$ close to $\pi$ are not
possible in the global fit (see upper panel of
Fig.~\ref{fig:chisq_delta} and the discussion in the following
subsection).

\begin{figure}[t] \centering 
    \includegraphics[width=0.44\textwidth]{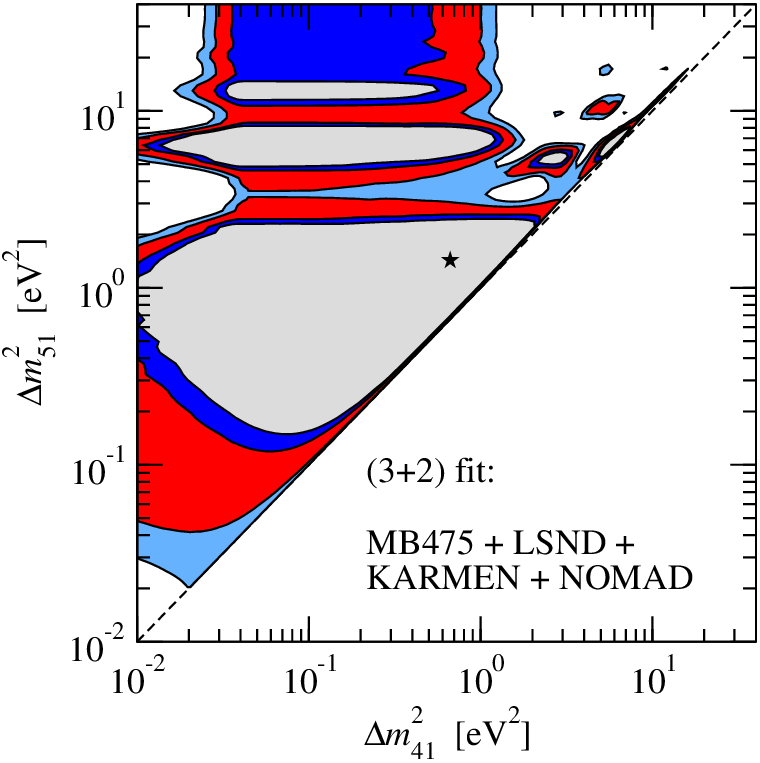}\hfil
    \includegraphics[width=0.44\textwidth]{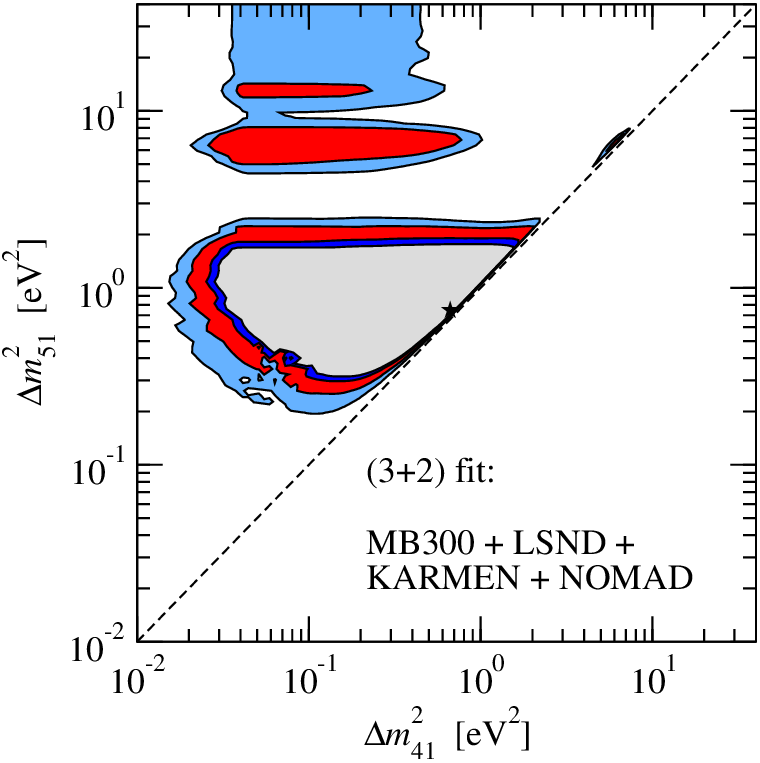}
    \mycaption{\label{fig:5nu-app}%
      Allowed regions for SBL appearance data in (3+2) schemes at
      90\%, 95\%, 99\%, 99.73\%~CL (2~dof) in the plane of $\Dmq_{41}$
      and $\Dmq_{51}$. All other parameters have been minimised. We
      use data from LSND, KARMEN, NOMAD, and MB475 (left) or MB300
      (right).}
\end{figure}

The allowed range in the plane of the two mass-squared differences is
shown in Fig.~\ref{fig:5nu-app}. Again we observe that the solution is
not particularly fine-tuned and a rather wide 90\%~CL region appears.
The fit using MB300 is somewhat more constrained by the requirement to
fit the low energy excess in MB.  The fact that for MB300 the best fit
occurs for $\Dmq_{41} \approx \Dmq_{51}$ is not statistically
significant; the allowed region extends far into the hierarchical
regime.

\subsection{Global SBL data in (3+2) schemes}

Now we proceed to the global analysis in (3+2) schemes, to see whether
the successful description of all appearance data found in the
previous sub-section can be reconciled also with the bounds from
disappearance experiments. The (3+2) survival probability in the SBL
approximation is given by
\begin{equation}
    P_{\nu_\alpha\to\nu_\alpha} = 1
    - 4\left(1 - \sum_{i=4,5} |U_{\alpha i}|^2 \right)
    \sum_{i=4,5} |U_{\alpha i}|^2 \, \sin^2\phi_{i1} 
    - 4\, |U_{\alpha 4}|^2|U_{\alpha 5}|^2 \, \sin^2\phi_{54}
\end{equation}
where $\phi_{ij}$ is given in Eq.~\eqref{eq:5nu-def}. Similar as in
the (3+1) case, also for (3+2) schemes atmospheric neutrino data
provide an important constraint on $\nu_\mu$ oscillations with sterile
neutrinos. The five-neutrino atmospheric neutrino analysis is
discussed in detail in the appendices~\ref{app:atm} and
\ref{app:robust}. It turns out that the same constraint
$\chi^2_\mathrm{ATM}(d_\mu)$ as in the four-neutrino case applies,
where now the definition $d_\mu = |U_{\mu 4}|^2 + |U_{\mu 5}|^2$ has
to be used (see appendix~\ref{app:atm}).

\begin{figure}[t] \centering 
    \includegraphics[width=0.95\textwidth]{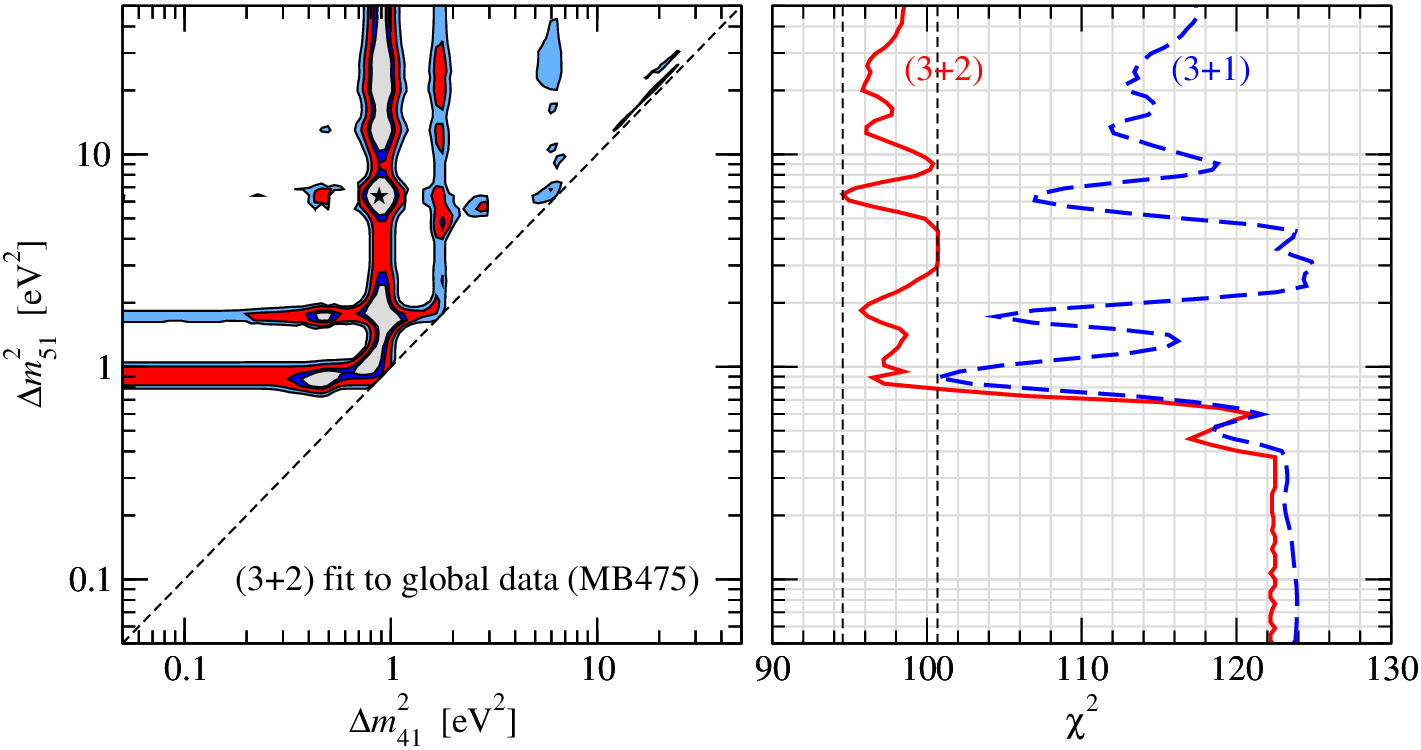}
    \mycaption{\label{fig:5nu-all_MB475}%
      Left: Allowed regions for global data in (3+2) schemes at 90\%,
      95\%, 99\%, 99.73\%~CL (2~dof) in the plane of $\Dmq_{41}$ and
      $\Dmq_{51}$, with all other parameters minimised. Right:
      $\chi^2$ projected onto the $\Dmq_{51}$ axis, using $\Dmq_{41}
      \le \Dmq_{51}$. For comparison we show also the $\chi^2$
      projection in (3+1) schemes. The vertical dashed lines indicate
      the $\chi^2$ of the (3+2) and (3+1) best fit points. For MB the
      two lowest energy data points have been omitted (MB475).}
\end{figure}

\begin{figure}[t] \centering 
    \includegraphics[width=0.95\textwidth]{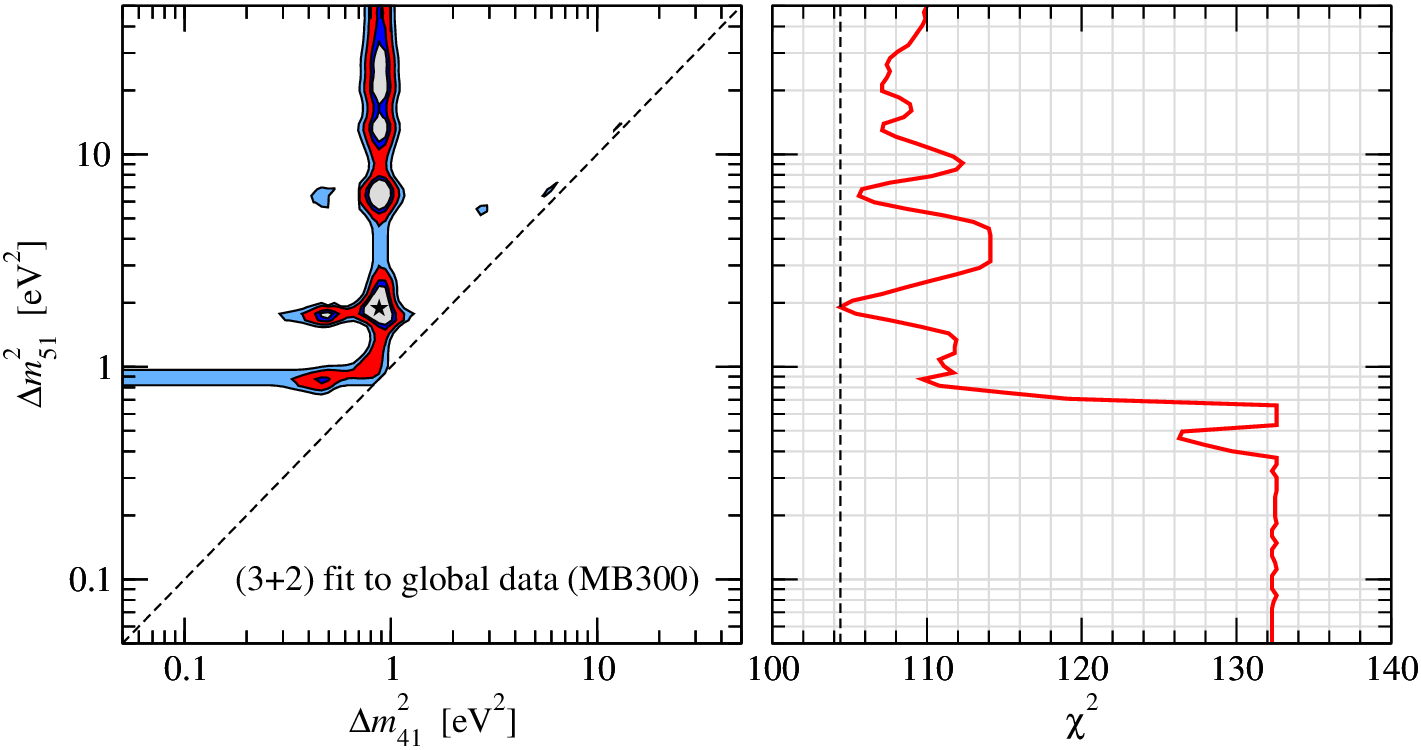}
    \mycaption{\label{fig:5nu-all_MB300}%
      Left: Allowed regions for global data in (3+2) schemes at 90\%,
      95\%, 99\%, 99.73\%~CL (2~dof) in the plane of $\Dmq_{41}$ and
      $\Dmq_{51}$, with all other parameters minimised. Right:
      $\chi^2$ projected onto the $\Dmq_{51}$ axis, using $\Dmq_{41}
      \le \Dmq_{51}$. The vertical dashed line indicates the $\chi^2$
      of the best fit point. For MB the two lowest energy data points
      have been included (MB300).}
\end{figure}

The results of our global (3+2) fit are summarised in the lower part
of Tab.~\ref{tab:bfp}, where the parameter values, the $\chi^2$ and
the gof of the best fit points are given, again for both MB options,
MB475 and MB300.  The allowed regions for the global fit in the plane
of the mass-squared differences are shown in
Figs.~\ref{fig:5nu-all_MB475} and \ref{fig:5nu-all_MB300}.

In the right panel of Fig.~\ref{fig:5nu-all_MB475} we show the
$\chi^2$ projections for (3+1) and (3+2) schemes. Comparing the two
best fit points provides a method to assess the relative quality of
the fit in the two models (likelihood ratio test).  We find that
introducing the second sterile neutrino leads to the relative
improvement of the fit of
\begin{equation} \label{eq:dchi}
    \chi^2_\text{min, global\,(3+1)} -
    \chi^2_\text{min, global\,(3+2)} = 6.1 \quad \text{(4 dof)} \,,
\end{equation}
where the number of dof corresponds to the additional 4 parameters
introduced by moving from (3+1) to (3+2). Hence, (3+1) can be rejected
only at the 81\%~CL with respect to the (3+2) model. This explains
also the ``stripes'' at $\Dmq_{41}$ and $\Dmq_{51}$ around $1~\eVq$,
which appear at 99\%~CL in Fig.~\ref{fig:5nu-all_MB475}. They
correspond to the (3+1) solution, which is always present as limiting
case in (3+2). Also note, that in the case of appearance data alone we
find
\begin{equation}
    \chi^2_\text{min, APP\,(3+1)} -
    \chi^2_\text{min, APP\,(3+2)} = 9.7 \quad \text{(3 dof)} \,.
\end{equation}
Comparing this number with Eq.~\eqref{eq:dchi} shows that the main
improvement in (3+2) schemes is to reconcile LSND and MB, whereas it
is not possible to evade efficiently the constraints from
disappearance data. This result is somewhat in disagreement with the
conclusion of Ref.~\cite{Sorel:2003hf}. A possible source of this
different result might be the inclusion of atmospheric neutrino data
in the fit, which is quite important to constrain sterile oscillations
in the $\nu_\mu$ sector. Our results are in accordance with the
arguments presented in the appendix of Ref.~\cite{Peres:2000ic}.

In the upper panel of Fig.~\ref{fig:chisq_delta} the $\chi^2$ of the
global fit is shown as a function of the complex phase $\delta$. One
can see from that figure that the global data prefer values of
$\delta$ close to ``maximal'' CP violation at $\delta = 3\pi/2$. The
best fit values are $1.64\pi$ and $1.44\pi$ for MB475 and MB300,
respectively. As discussed in appendix~\ref{app:CP}, reconciling MB
and LSND with values of $\delta$ very close $\pi$ (as found for
appearance data only) requires rather large values of the appearance
amplitudes $|U_{e4} U_{\mu 4}|$ and $|U_{e5} U_{\mu 5}|$, close to the
upper bound from unitarity. Such large values are not compatible
with disappearance data, and hence solutions with $\delta$ close to
$\pi$ are not possible in the global fit.

From the $\chi^2$ values given in Tab.~\ref{tab:bfp} it appears that
the model provides a very good fit to the data.  However, as in the
(3+1) case the problem appears when the compatibility of different
data sets is considered.
Let us divide the global data into appearance and disappearance
experiments and check their compatibility with the PG
test~\cite{Maltoni:2003cu} according to Eq.~\eqref{eq:PG}. We find the
following $\chi^2_\text{PG}$ values, for global data without MB, with
MB475, and with MB300:\footnote{We have tested this result explicitly
with the ``official'' MB analysis available at Ref.~\cite{MB-data}.
Using the MB $\chi^2$ from that source the PG test for appearance and
disappearance data gives $\chi^2_\mathrm{PG} = 17.9$ (MB475) and 24.6
(MB300), in good agreement with our results displayed in
Eq.~\eqref{eq:5nuPG}.}
\begin{equation} \begin{aligned} \label{eq:5nuPG}
    \chi^2_\text{PG} &= 17.5\,, \quad
    &\text{PG} &= 1.5\times 10^{-3} \qquad \text{(no MB)}
    \\
    \mbox{APP vs DIS:} \qquad
    \chi^2_\text{PG} &= 17.2 \,, \quad 
    & \text{PG} &= 1.8\times 10^{-3} \qquad \text{(MB475)}
    \\
    \chi^2_\text{PG} &= 25.1 \,, \quad
    & \text{PG} &= 4.8\times 10^{-5} \qquad \text{(MB300)}
\end{aligned} \end{equation}
The PG values have been calculated for 4~dof~\cite{Maltoni:2003cu}.
This number corresponds to the four independent parameter
(combinations) $|U_{e4} U_{\mu 4}|,|U_{e5} U_{\mu 5}|, \Delta
m^2_{41}, \Delta m^2_{51}$, representing the minimal number of
parameter (combinations) in common to the two data sets. 
From Eq.~\eqref{eq:5nuPG} we conclude that also in (3+2) schemes a
severe tension exists between appearance and disappearance
experiments. If MB475 is used the result is very similar to the
situation without MB data implying inconsistency at about $3.1\sigma$,
whereas in case of the full MB data the tension becomes significantly
worse (about $4\sigma$), since appearance data are more constraining
because of the need to accommodate LSND as well as the MB excess at
low energies.

\begin{figure}[t] \centering 
    \includegraphics[width=0.9\textwidth]{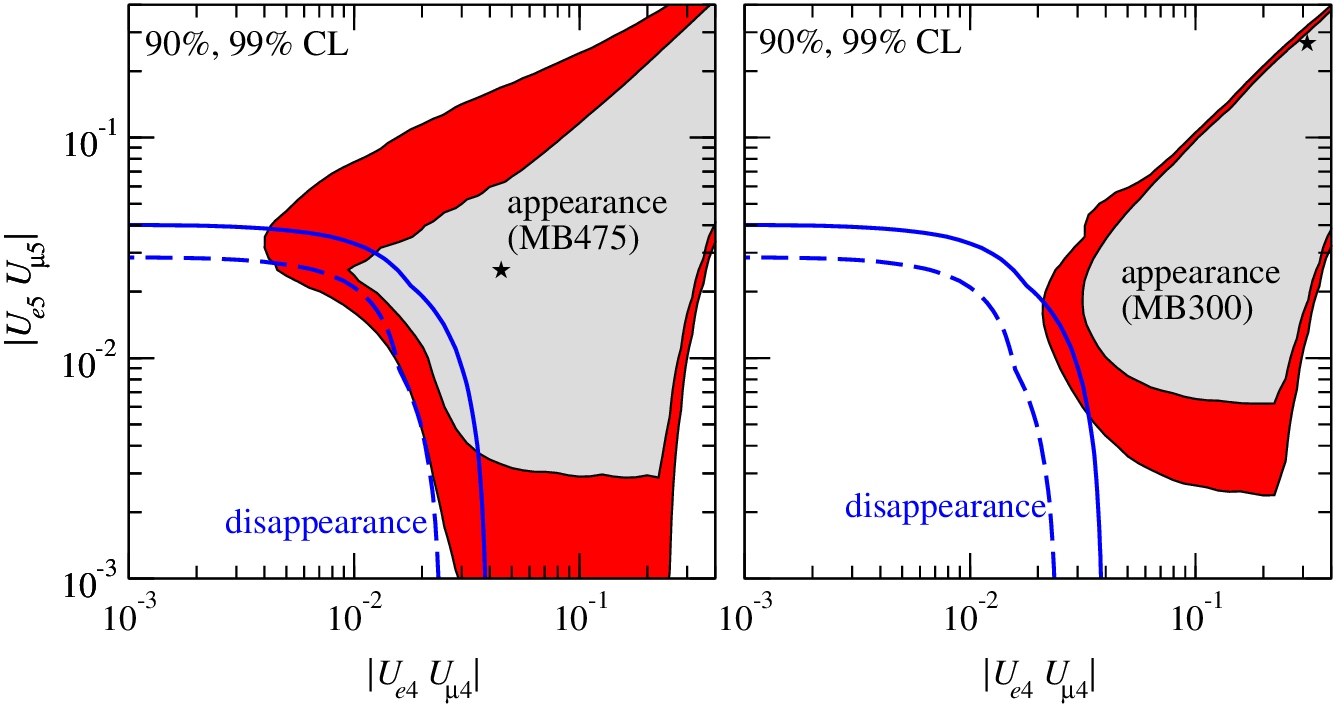}
    \mycaption{\label{fig:app-vs-dis}%
      Allowed regions at 90\% and 99\%~CL in (3+2) schemes for
      appearance data (shaded regions) and disappearance data (dashed
      and solid curves) projected onto the plane of $|U_{e4} U_{\mu
      4}|$ and $|U_{e5} U_{\mu 5}|$. In the left panel the two lowest
      energy data points in MB have been omitted (MB475), whereas in
      the right panel the full MB energy range has been used in the
      fit (MB300).}
\end{figure}

The tension between appearance and disappearance data is illustrated
in Fig.~\ref{fig:app-vs-dis}, where we show the projections of the
allowed regions in the plane of the appearance amplitudes $|U_{e4}
U_{\mu 4}|$ and $|U_{e5} U_{\mu 5}|$.  The opposite trend of the two
data sets is clearly visible, especially when the low energy excess in
MB is included (right panel). Note that an overlap of the regions
visible in that figure does not prove that there is indeed an overlap
of the allowed regions in the full parameters space since only a
projection is shown. The ``common'' values in the plane shown in the
plot might correspond actually to different locations in the space of
$\Dmq_{41}$ and $\Dmq_{51}$. However, if no overlap is visible in that
projection at a certain CL there is also no overlap at that CL in the
full parameter space.

Comparing the numbers for MB475 and MB300 given in
Eq.~(\ref{eq:5nuPG}) it becomes obvious that the MB low energy excess
is a severe problem in the global (3+2) fit, although a very good fit
can be obtained for appearance data only. This is also apparent from
the $\chi^2_\mathrm{min}$ values given in Tab.~\ref{tab:bfp}: Adding
the two additional MB data points at low energy leads to an increase
of the best fit $\chi^2$ of about 10 units from 94.5 to 104.4. Indeed,
using the global data the MB excess cannot be fitted, as visible in
the right panel of Fig.~\ref{fig:spectrum}, where we show the
prediction for the MB spectrum at the global best fit point. The
reason is that to explain the excess relatively large values of
$|U_{e4} U_{\mu 4}|$ and $|U_{e5} U_{\mu 5}|$ are required (see
Fig.~\ref{fig:app-vs-dis}, right), which are inconsistent with
disappearance data.

Before closing this section we give the results of an alternative
consistency test for the (3+2) model. Instead of dividing the global
data into appearance and disappearance experiments, we now consider
the two data sets LSND and all the remaining NEV data, similar as done
in Eq.~\eqref{eq:PG4nu} and Fig.~\ref{fig:4nu-bound} for the (3+1)
schemes. In case of (3+2) this analysis gives
\begin{equation}
    \chi^2_\text{PG} = 21.2 \,(5\,{\rm dof}), \quad
    \text{PG} = 7.8\times 10^{-4} \, (3.4\sigma) 
    \qquad \text{LSND vs NEV (incl. MB475)} \,.
\end{equation}
Here 5 dof have been used, corresponding to the 5 parameters in
common.  Without MB we find in this case $\chi^2_\text{PG} = 19.6$,
$\text{PG} = 1.5\times 10^{-3} \, (3.2\sigma)$.  Hence, the PG test
gives a disagreement between LSND and the remaining SBL data 
similar to the disagreement between appearance and disappearance data
found in Eq.~\eqref{eq:5nuPG}.


\section{(3+3) six-neutrino mass schemes}
\label{sec:3+3}

Since there are three active neutrinos it seems natural to consider
also the case of three sterile neutrinos. If all three additional
neutrino states have masses in the eV range and mixings as relevant
for the SBL experiments under consideration, such a model will
certainly have severe difficulties to accommodate standard
cosmology~\cite{cosmo-bounds}, and one has to refer to some
non-standard cosmological scenario~\cite{L-asymm, BBN-majoron,
Gelmini:2004ah}. Here we leave this problem aside and focus on
neutrino oscillation data, investigating how much the fit improves
with respect to the five-neutrino case. 

The relevant oscillation probabilities are easily generalised to the
(3+3) scheme:
\begin{multline} \label{eq:6nu-prob}
    P_{\nu_\mu\to\nu_e} =
    4 \, \sum_i |U_{ei}|^2 |U_{\mu i}|^2 \, \sin^2 \phi_{i1} \\
 +  8 \, \sum_{i, j < i} |U_{ei}U_{\mu i}U_{ej}U_{\mu j}| \,
    \sin\phi_{i1}\sin\phi_{j1}\cos(\phi_{ij} - \delta_{ij}) \,,
    \qquad i, j = 4,5,6\,,
\end{multline}
with the definitions
\begin{equation} \label{eq:6nu-def}
    \phi_{ij} \equiv \frac{\Dmq_{ij}L}{4E} \,,
    \qquad \delta_{ij} \equiv
    \arg\left(U_{ej}^* U_{\mu j} U_{ei} U_{\mu i}^* \right) \,. 
\end{equation}
Eq.~\eqref{eq:6nu-prob} holds for neutrinos, for anti-neutrinos one
has to replace $\delta_{ij} \to -\delta_{ij}$. Note that only two
phases $\delta_{ij}$ are independent. The survival probability is
given by 
\begin{multline}
    P_{\nu_\alpha\to\nu_\alpha} = 1
    - 4\left(1 - \sum_i |U_{\alpha i}|^2 \right)
    \sum_i |U_{\alpha i}|^2 \, \sin^2\phi_{i1} \\
    - 4\, \sum_{i, j<i} |U_{\alpha i}|^2|U_{\alpha j}|^2 \, \sin^2\phi_{ij}
    \,, \qquad i, j = 4,5,6 \,.
\end{multline}
Atmospheric data is included in a similar way as in the previous
cases, by using $\chi^2(d_\mu)$, where now we define $d_\mu =
\sum_{i=4,5,6} |U_{\mu i}|^2$, see appendix~\ref{app:atm}.  In
general, one of the new mass-splittings could also fall into the
atmospheric range of few$\times 10^{-3}$~eV$^2$. We have not
considered such degenerate situations, and we always assume that all
$\Dmq_{i1}$, $\Dmq_{ij}$, with $i,j=4,5,6$ are infinite for the
atmospheric neutrino analysis. 

We have scanned the $\chi^2$ of global data in the 11 dimensional
parameter space, where a grid of $81\times 81 \times 81$ values for
the mass-squared differences has been used, spaced logarithmically
from 0.1 to 20~eV$^2$. In each point of this grid the remaining 8
parameters have been minimised by a standard optimisation routine.

\begin{figure}[t] \centering 
    \includegraphics[width=0.9\textwidth]{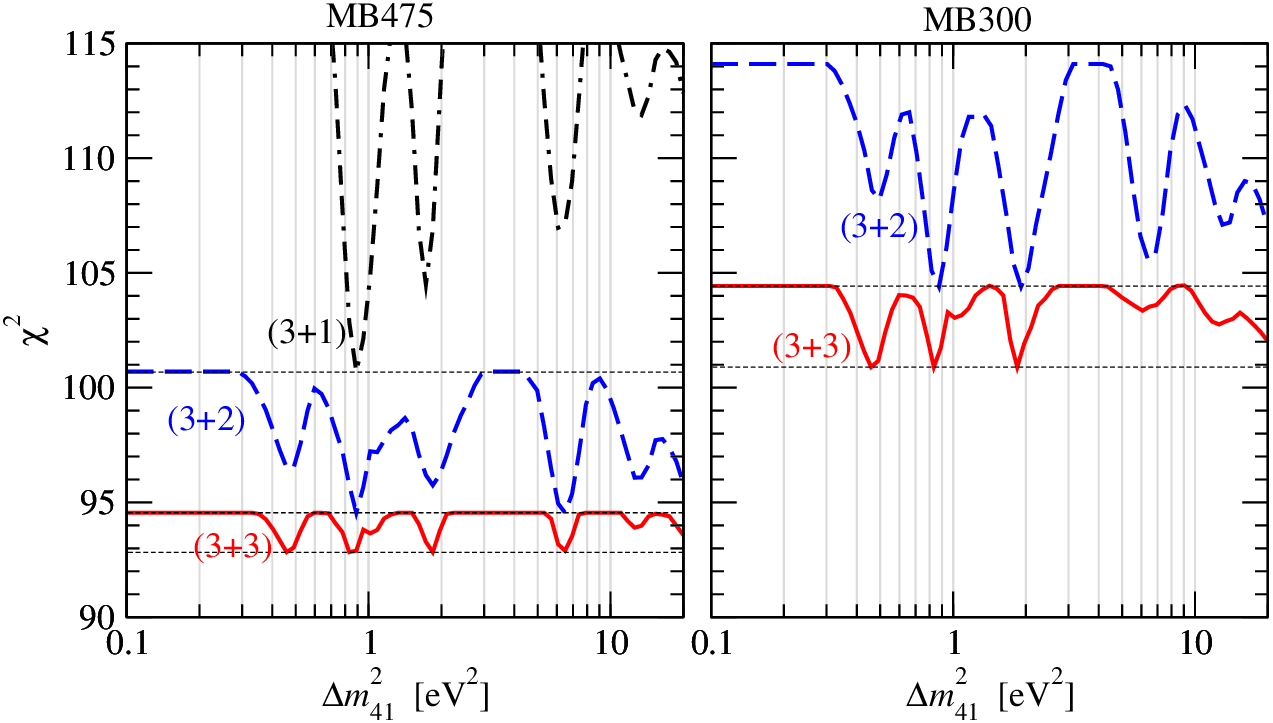}
    \mycaption{\label{fig:6nu}%
      $\chi^2$ of global data in (3+3) six-neutrino mass schemes
      projected onto the $\Delta m^2_{41}$ axis, using MB475 (left) or
      MB300 (right). For comparison also the $\chi^2$ is shown for the
      (3+2) and (3+1) schemes. The horizontal dashed lines indicate
      the corresponding best fit $\chi^2$ values.}
\end{figure}

The results of our search are shown in Fig.~\ref{fig:6nu}, where we
display the global $\chi^2$ as a function of $\Dmq_{41}$. For
comparison also the $\chi^2$ curves in the case of (3+2) and for MB475
also for (3+1) are shown. Note that different from Sec.~\ref{sec:3+2},
here we do not impose any constraint on the ordering of the
mass-squared differences, and the full $s$-dimensional space is
scanned for the (3+$s$) scheme. This implies that the result is
symmetric with respect to the $s$ mass-squared differences, and the
$\chi^2$ functions projected onto any of the $(\Dmq_{i1})$-axes ($i
\ge 4$) are identical. (Just for convenience we label the horizontal
axis in Fig.~\ref{fig:6nu} as $\Dmq_{41}$, since it is available in
all schemes.) Furthermore, as a consequence of this symmetry there are
2 (3) degenerate minima in the five (six) neutrino case, which
corresponds to a re-labelling of the mass states. Note also, as it must
be, the $\chi^2_\mathrm{min}$ of the (3+($s-1$)) model is the maximal
$\chi^2$ value in (3+$s$), since the (3+($s-1$)) fit is always
available as limiting case.

\begin{table}[t] \centering
    \begin{tabular}{l@{\quad}c@{\quad}c@{\quad}c@{\quad}c@{\quad}c@{\quad}c@{\quad}c}
	\hline\hline
	& $\Dmq_{41}$ & $\Dmq_{51}$  & $\Dmq_{61}$ 
        & $\chi^2_\text{min} / \text{dof}$ & gof 
        & $\chi^2_{(3+2)} - \chi^2_{(3+3)}$ & CL
	\\
	\hline
	MB475 & 0.46 & 0.83 & 1.84 & $92.8/(107-11)$ & 57\% & $1.7/4$ & 20\% 
        \\
	MB300 & 0.46 & 0.83 & 1.84 & $100.9/(109-11)$& 40\% & $3.5/4$ & 52\%
	\\
	\hline\hline
    \end{tabular}
    \mycaption{\label{tab:6nubfp}%
      Best fit points for global data in the (3+3) scheme.
      Mass-squared differences are given in eV$^2$. We give also the
      $\chi^2$ difference between the (3+2) and (3+3) best fits. The
      last column shows the CL at which (3+2) is accepted with respect
      to (3+3), as derived from the $\chi^2$ difference evaluated for
      4~dof, corresponding to four additional parameters in the (3+3)
      model.}
\end{table}

The global (3+3) best fit points are summarised in
Tab.~\ref{tab:6nubfp}. From the table and Fig.~\ref{fig:6nu} one can
see that there is only a marginal improvement of the fit by 1.7 units
in $\chi^2$ for MB475 (3.5 for MB300) with respect to (3+2), to be
compared with 4 additional parameters in the model. Hence, we conclude
that there are no qualitatively new effects in the (3+3) scheme. The
conflict between appearance and disappearance data remains a problem,
and the additional freedom introduced by four new parameters does not
relax significantly this tension.


\section{Summary}
\label{sec:summary}

We have considered the global fit to SBL neutrino oscillation data including
the recent data from the MiniBooNE (MB) experiment~\cite{MB} in the framework
of four-, five-, and six-neutrino oscillations. We have divided the global
data into various sub-sets and tested their consistency within the
sterile-neutrino oscillation framework. These results are summarised in
Tab.~\ref{tab:summary} for the (3+1) and (3+2) schemes. Clearly, in all cases
we find severe tension between different sub-samples of the data, with the
only exception when LSND and the low-energy excess in MB are left out, and in
this case indeed no sterile neutrinos are needed and the standard three active
neutrino scheme (3+0) provides a perfect fit to the data.

\begin{table}
  \begin{tabular}{l@{\quad}ccc@{\quad}ccc}
  \hline\hline
  & \multicolumn{3}{c}{(3+1)} & \multicolumn{3}{c}{(3+2)} \\
  \hline
  & $\chi^2_\mathrm{global}$/dof & $\chi^2_\mathrm{PG}$/dof & PG &
    $\chi^2_\mathrm{global}$/dof & $\chi^2_\mathrm{PG}$/dof & PG \\
  \hline
  DIS vs K+N+L       &  95.5/96  & 14.8/2 & $6.1\times 10^{-4}$
                     & 92.1/92  & 17.4/4 & $1.5\times 10^{-3}$ \\
  DIS vs K+N+L+MB475 & &&& 94.5/100  & 17.2/4 & $1.8\times 10^{-3}$ \\
  DIS vs K+N+L+MB300 & &&& 104.4/102 & 25.1/4 & $4.8\times 10^{-5}$ \\
  DIS vs K+N+MB475   &  70.5/93  & 0.1/2  & 0.95 
                     &  68.9/89  & 1.1/4  & 0.89 \\
  DIS vs K+N+MB300   & &&& 79.1/91   & 10.3/4 & $3.6\times 10^{-2}$ \\
  NEV vs L           &  95.5/96  & 20.9/2 & $2.9\times 10^{-5}$ 
                     &  92.1/92  & 19.6/5 & $1.5\times 10^{-3}$ \\
  NEV+MB475 vs L     & 100.7/104 & 24.7/2 & $4.3\times 10^{-6}$ 
                     &  94.5/100 & 21.2/5 & $7.8\times 10^{-4}$ \\
  \hline\hline
  \end{tabular}
  \mycaption{\label{tab:summary} Summary table for various consistency
  checks within the (3+1) and (3+2) schemes. The PG test
  \cite{Maltoni:2003cu} as been defined in Eq.~(\ref{eq:PG}). In the
  first column, where we give the data sets tested against each other,
  we use the following abbreviations: K (KARMEN), N (NOMAD), L (LSND),
  DIS corresponds to the disappearance experiments summarised in
  Eq.~(\ref{eq:data}), and NEV=DIS+K+N. Results for the (3+3) scheme are
  qualitatively similar to (3+2).}
\end{table}

Let us summarise our findings:
\begin{enumerate}
  \item
    {\bf (3+1)} four-neutrino schemes are strongly disfavoured because
    \begin{enumerate}
      \item
	recent MB data is incompatible with LSND at the
	98\%~CL~\cite{MB},
      \item
	the tension between LSND and NEV SBL data becomes more severe
	due to MB, there is no overlap of the allowed regions for NEV
	and LSND at 99\%~CL, and the PG test implies inconsistency at
	the level of $4\sigma$,
      \item
	it is not possible to account for the low energy event excess
	in MB.
    \end{enumerate}
  \item
    {\bf (3+2)} five-neutrino schemes 
    \begin{enumerate}
      \item
	do provide a good fit to LSND and the recent MB data,
      \item
	they can account for the low energy event excess in MB,
	however
      \item
	there is significant tension between appearance and
	disappearance data (according to the PG test at the level of
	$3\sigma$ for MB475 and $4\sigma$ for MB300).
    \end{enumerate}
  \item
    {\bf (3+3)} six-neutrino schemes do not offer qualitatively new
    effects, the global $\chi^2$ improves only by about 1.7 (3.5)
    units for MB475 (MB300) with respect to (3+2), and hence, the
    conflict between appearance and disappearance data remains.
\end{enumerate}
The points 2a and 2b might be considered as an interesting hint in
favour of (3+2) schemes. Since the combined fit of LSND and MB is based
on a non-trivial complex phase which introduces a difference in
neutrino and anti-neutrino oscillations, these results would represent
the first indication of CP violation in neutrino oscillations. This
hypothesis could be tested by MB anti-neutrino data, which is
currently being accumulated. However, point 2c is a challenge for the
(3+2) model. The conclusions of 2c and 3 strongly rely on the
disappearance experiments Bugey and CDHS. A crucial check would be the
confirmation of $\nu_e$ or $\nu_\mu$ disappearance at the $1~\eVq$
scale. Hence, it might be worth to investigate the possibility to
obtain such information at future reactor experiments~\cite{reactor},
from disappearance experiments based on low-energy neutrinos from
radio-active sources~\cite{low-energy}, or at the near detector
complex of up-coming long-baseline superbeam experiments~\cite{LBL}.
A characteristic signal of sterile neutrino oscillations could be
obtained at experiments exploring neutral-current
detection~\cite{Garvey:2005pn}.


\acknowledgments

We thank Michel Sorel for communication on the MiniBooNE experiment
and useful comments on our analysis. MM is supported by MCYT through
the Ram\'on y Cajal program, by CiCYT through the project
FPA2006-01105 and by the Comunidad Aut\'onoma de Madrid through the
project P-ESP-00346.


\appendix

\section{Reconciling LSND and MB in (3+2) schemes}
\label{app:CP}

\begin{figure}[t] \centering 
    \includegraphics[width=0.88\textwidth]{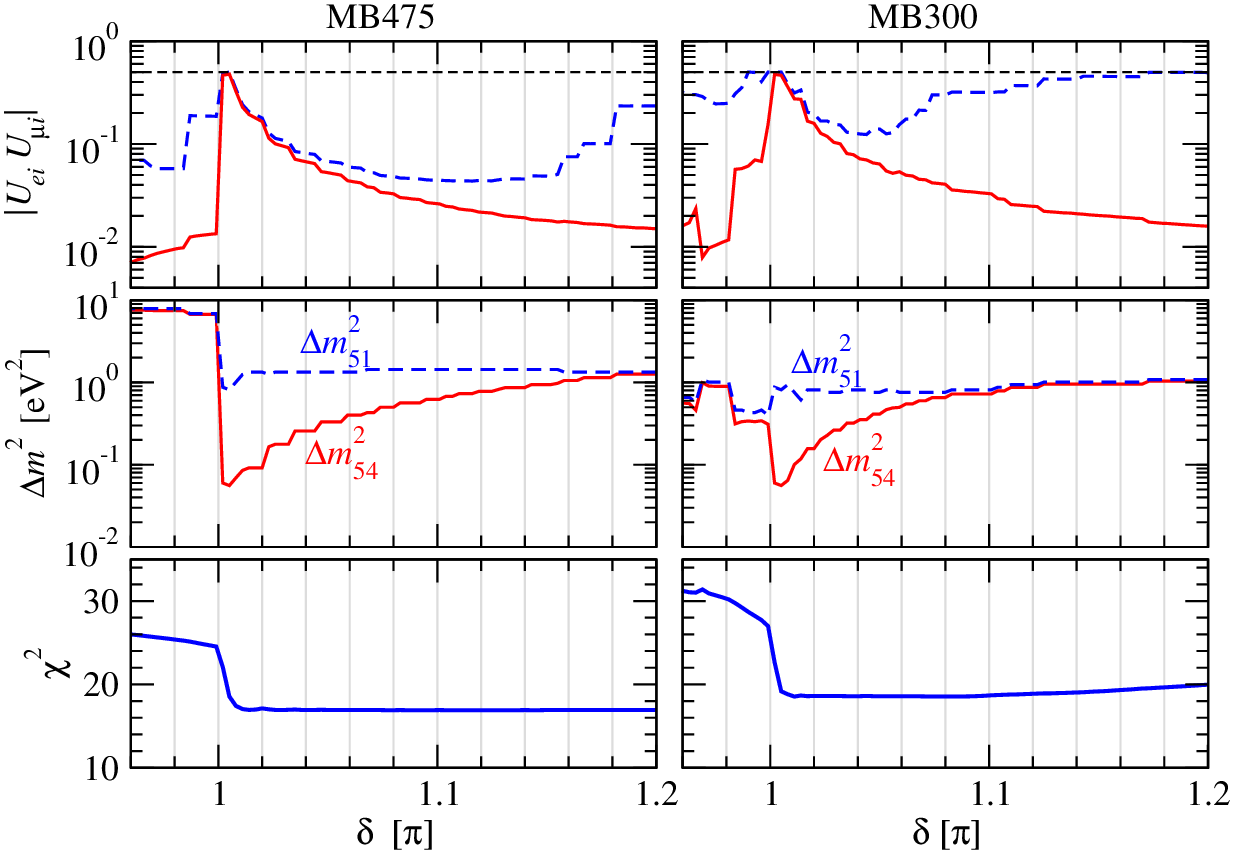}
    \mycaption{\label{fig:CP}%
      Fit to appearance data LSND, KARMEN, NOMAD, and MB475 (left) or
      MB300 (right) in (3+2) schemes. We show the $\chi^2$ (bottom),
      the values of $\Dmq_{51}$, $\Dmq_{54}$ (middle), and the values
      of $q_i \equiv |U_{ei} U_{\mu i}|$, $i=4,5$ (top) as a function
      of the complex phase $\delta$ defined in Eq.~\ref{eq:5nu-def}.
      The horizontal dashed lines in the top panels correspond to the
      maximal value allowed by unitarity of the mixing matrix.}
\end{figure}

In this appendix we discuss in some detail how LSND and MB are
reconciled in (3+2) schemes exploring CP violation in the appearance
probability. In particular, it is intriguing that a very good fit can
be obtained with a complex phase $\delta$ very close to the CP
conserving value $\pi$, compare Fig.~\ref{fig:chisq_delta}. To
understand this effect we show in Fig.~\ref{fig:CP} a zoom into the
region around $\delta = \pi$, and we display in addition to the
$\chi^2$ also the values obtained for the oscillation parameters.

Let us consider the probability $P_{\nu_\mu\to\nu_e}$ given in
Eq.~\eqref{eq:5nu-prob}. A non-trivial possibility to suppress this
probability can be obtained by requiring $\cos(\phi_{54} - \delta) =
-1$. Then one has 
\begin{equation}\label{eq:P1}
P_{\nu_\mu\to\nu_e} = 4 (q_4 \sin\phi_{41} - q_5 \sin\phi_{51})^2 \,,
\qquad \cos(\phi_{54} - \delta) = -1 \,,
\end{equation}
with the abbreviation $q_i \equiv |U_{ei} U_{\mu i}|$. Hence, the
probability is small for $q_4 \approx q_5$ and $\phi_{54} \ll 1$.
This is precisely the behaviour shown in Fig.~\ref{fig:CP}: when
$\delta$ approaches $\pi$ from above, $\Dmq_{54}$ becomes small and
the $q_i$ approach each other. Writing $\delta = \pi + \epsilon$ one
has $\cos(\phi_{54} - \delta) \approx -1 +
\mathcal{O}(\phi_{54}^2,\epsilon^2)$, Eq.~\eqref{eq:P1} is valid, and
the oscillation probability is suppressed in MB. 

Now the question arises whether large enough values for
$P_{\bar\nu_\mu\to\bar\nu_e}$ can be achieved in order to explain
LSND. The difference of anti-neutrino and neutrino probabilities is
given by
\begin{equation}\label{eq:DP}
P_{\bar\nu_\mu\to\bar\nu_e} - P_{\nu_\mu\to\nu_e} =
16 \, q_4q_5 \, \sin\phi_{41} \, \sin\phi_{51} \, 
\sin\phi_{54} \, \sin\epsilon 
\approx
16 \, q_4q_5 \, \sin^2(\phi_{51}) \, \phi_{54} \epsilon \,,
\end{equation}
where in the last step $\phi_{54},\epsilon \ll 1$ has been used. Since
$\phi_{54}$ and $\epsilon$ are small, the other factors have to be as
large as possible in order to get a sufficient probability for LSND.
Indeed, for $\Dmq_{51} \approx 1$~eV$^2$ one has $\sin^2\phi_{51}
\approx 1$, and also the $q_i$ grow for $\epsilon \to 0$ (see
Fig.~\ref{fig:CP}). Once the maximal values allowed by unitarity, $q_4
= q_5 = 1/2$, are reached the LSND probability is given roughly by
$P_{\bar\nu_\mu\to\bar\nu_e} \sim 4 \epsilon^2$, where we used
$P_{\nu_\mu\to\nu_e} \approx 0$ (in order to explain MB) and
$\phi_{54} \approx \epsilon$ (in order to have $\cos(\phi_{54} -
\delta) \approx -1$). Using the experimental value $P_\mathrm{LSND} =
0.0026$ one finds that a fit should be possible for $\epsilon \gtrsim
0.025 \approx 0.008\pi$, in agreement with our results.

The similar structure of the left and right panels of
Fig.~\ref{fig:CP} suggests that this mechanism works equally well for
MB475 and MB300, and fitting the low energy excess in MB does not
affect these considerations. 
Obviously, this explanation is not valid for $\delta < \pi$, since the
CP asymmetry Eq.~\eqref{eq:DP} has the wrong sign to reconcile LSND
and MB. As visible in Fig.~\ref{fig:CP}, the fit jumps into a quite
different solution, which anyway gives a poor $\chi^2$.
Also, the local minimum around $\delta \sim \pi/2$ visible in
Fig.~\ref{fig:chisq_delta} for MB475 requires a different explanation
in order to obtain the correct sign of the CP asymmetry for these
values of $\delta$.
Let us also mention that quite large values of $q_4$ and $q_5$ close
to the unitarity bound do appear in the fit for $\delta \gtrsim \pi$,
since only appearance experiments are used. Such large values are not
possible if disappearance experiments are included, which basically
require that each of the $|U_{ei}|$, $|U_{\mu i}|$, $i=4,5$ has to be
small. This is one reason for the difficulties in reconciling
appearance and disappearance data, in close analogy to (3+1).


\section{Oscillations with extra sterile states}
\label{app:atm}

In this appendix we discuss in some detail atmospheric and
short-baseline neutrino oscillations involving extra sterile neutrino
states. For definiteness, we will focus on (3+3) schemes; expressions
for (3+2) and (3+1) models can be easily obtained by dropping all
terms containing a redundant ``6'' or ``5'' index. Let us order the
flavor eigenstates as $(\nu_e,\, \nu_\mu,\, \nu_\tau,\, \nu_{s_1},\,
\nu_{s_2},\, \nu_{s_3})$ and introduce the following parametrisation
for the neutrino mixing:
\begin{equation} \label{eq:Uglob}
    U = \tilde{R}_{36} \, \tilde{R}_{35} \, \tilde{R}_{34} \,
    R_{26} \, R_{25} \, R_{24} \, R_{23} \, \tilde{R}_{16} \,
    \tilde{R}_{15} \, R_{14} \, \tilde{R}_{13} \, \tilde{R}_{12} \,,
\end{equation}
where $\tilde{R}_{ij}$ represents a complex rotation by an angle
$\theta_{ij}$ and a phase $\varphi_{ij}$ in the $ij$ plane, while
$R_{ij}$ is an ordinary rotation by an angle $\theta_{ij}$.
Note that rotations involving only sterile states (\textit{i.e.},
$\tilde{R}_{\ell\ell'}$ with both $\ell, \ell' \ge 4$) are unphysical,
and therefore we have omitted them from Eq.~\eqref{eq:Uglob}.
For the general case with $s$ sterile states, it is convenient to
choose the $(s+2)$ rotations $R_{14}$ and $R_{2j}$ as real, and the
remaining ones as complex. The matrix $U$ then includes $3(s+1)$
angles and $(2s+1)$ phases.

A number of simplifying assumptions can be made in the analysis of
short-baseline as well as atmospheric and long-baseline neutrino
experiments.
For short-baseline experiments one can neglect the solar and
atmospheric mass splittings, $\Dmq_{21} = 0$ and $\Dmq_{31} = 0$. In
this approximation, the mixing angles $\tilde\theta_{12}$,
$\tilde\theta_{13}$ and $\theta_{23}$ disappear from the relevant
probabilities. Furthermore, matter effects can be neglected. Since we
do not consider neutral current interactions in our analysis, the
$\tau$ neutrino is essentially indistinguishable from the sterile
states, as it participates neither in the production nor in the
detection processes. Therefore, all the angles $\tilde\theta_{3j}$
also disappear. So for (3+3) models we are left with an effective
mixing matrix
\begin{equation}\label{eq:Usbl}
    U = R_{26} \, R_{25} \, R_{24} \, \tilde{R}_{16} \, \tilde{R}_{15}
    \, R_{14} \qquad \text{for SBL},
\end{equation}
which contains six angles and two CP phases. In general, under our
approximations SBL experiments depend on $2s$ angles and $(s-1)$
phases. For example, in (3+2) models we have four angles
($\theta_{14}$, $\theta_{15}$, $\theta_{24}$, $\theta_{25}$) and one
phase ($\varphi_{15}$), and the matrix elements $|U_{e4}|$,
$|U_{e5}|$, $|U_{\mu 4}|$, $|U_{\mu 5}|$, $\arg(U_{e4}^* U_{\mu 4}
U_{e5} U_{\mu 5}^*)$ used in Sec.~\ref{sec:3+2} are combinations of
these five parameters.

For atmospheric and long-baseline experiments (K2K and MINOS) we
neglect the mixing of $\nu_e$ with other neutrino states at the LSND
mass-squared splittings, justified by the constraint from Bugey. This
corresponds to setting all the angles $\tilde\theta_{1\ell}$ with
$\ell \ge 4$ to zero. In this approximation, the complex phase
$\varphi_{12}$ can be dropped.  Therefore, in (3+3) models we are left
with an effective mixing matrix
\begin{equation} \label{eq:Uatm}
    U = \tilde{R}_{36} \, \tilde{R}_{35} \, \tilde{R}_{34} \, R_{26}
    \, R_{25} \, R_{24} \, R_{23} \, \tilde{R}_{13} \, R_{12}
    \qquad \text{for ATM and LBL},
\end{equation}
which contains nine angles and four CP phases. As a general rule, in
our approximation for ATM and LBL experiments the matrix $U$ contains
$(2s+3)$ angles and $(s+1)$ phases.

From Eqs.~\eqref{eq:Usbl} and~\eqref{eq:Uatm} it is straightforward to
see that for any number of extra sterile states, $s$, atmospheric and
short-baseline experiments are connected only through the $s$ angles
$\theta_{2\ell}$ with $\ell \ge 4$ (or, equivalently, through the
parameters $|U_{\mu\ell}|^2$ with $\ell \ge 4$). Note that in our
convention all the non-vanishing CP phases are ``private'' to either
short-baseline (\textit{e.g.}, $\varphi_{15}$ and $\varphi_{16}$) or
atmospheric (\textit{e.g.}, $\varphi_{13}$, $\varphi_{34}$,
$\varphi_{35}$ and $\varphi_{36}$) experiments.

\bigskip

Let us now focus on the probabilities relevant for the analysis of
atmospheric and long-baseline experiments. The Hamiltonian in the
flavor basis is:
\begin{equation}
    H = U \Delta U^\dagger + V \,,
\end{equation}
where $U$ is given in Eq.~\eqref{eq:Uatm}, $\Delta = \diag(0,\,
\Dmq_{21},\, \Dmq_{31},\, \Dmq_{41},\, \Dmq_{51},\, \dots) / 2E$, and
$V = \pm \sqrt{2} \, G_F \diag(2 N_e,\, 0,\, 0,\, N_n,\, N_n,\,
\dots)/2$. It is convenient to define $U_\text{SM} = R_{23} \,
\tilde{R}_{13} \, R_{12}$ and $\tilde{U} = U \, U_\text{SM}^\dagger$.
Then we can write:
\begin{equation} \label{eq:Htilde}
    H = \tilde{U} \tilde{H} \tilde{U}^\dagger
    \qquad\text{with}\qquad
    \tilde{H} = U_\text{SM} \Delta U_\text{SM}^\dagger
    + \tilde{U}^\dagger V \tilde{U} \,.
\end{equation}
In order to further simplify the analysis, let us now assume that all
the mass-squared differences involving the ``heavy'' states $\nu_\ell$
with $\ell \ge 4$ can be considered as infinite: $\Dmq_{\ell i},
\Dmq_{\ell\ell'} \to \infty$ for any $i=1,2,3$ and $\ell,\ell' \ge 4$.
In leading order, the matrix $\tilde{H}$ takes the effective
block-diagonal form:
\begin{equation}
    \tilde{H} \approx
    \begin{pmatrix}
	H^{(3)} & \boldsymbol{0} \\
	\boldsymbol{0} & \Delta^{(s)}
    \end{pmatrix}
\end{equation}
where $H^{(3)}$ is the $3\times 3$ sub-matrix of $\tilde{H}$
corresponding to the first, second and third neutrino states, and
$\Delta^{(s)} = \diag(\Dmq_{41},\, \Dmq_{51},\, \dots)/ 2E$ is a
diagonal $s\times s$ matrix (the matter terms in this block are
negligible in the limit of very large $\Dmq_{\ell\ell'}$).
Consequently, the evolution matrix is:
\begin{equation}
    \tilde{S} \approx
    \begin{pmatrix}
	e^{i H^{(3)} L} & \boldsymbol{0} \\
	\boldsymbol{0} & e^{i \Delta^{(s)} L}
    \end{pmatrix}
    \qquad\text{and}\qquad
    S = \tilde{U} \tilde{S} \tilde{U}^\dagger \,.
\end{equation}
We are interested only in the elements $S_{ee}$, $S_{e\mu}$, $S_{\mu
e}$ and $S_{\mu\mu}$. Taking into account the block-diagonal form of
$\tilde{S}$ and the relations $\tilde{U}_{ei} = \delta_{i1}$ and
$\tilde{U}_{\mu 1} = \tilde{U}_{\mu 3} = 0$, we obtain:
\begin{equation}
    S_{ee} = \tilde{S}_{11} \,,
    \quad
    S_{e\mu} = \tilde{U}_{\mu 2}^\star \, \tilde{S}_{12} \,,
    \quad
    S_{\mu e} = \tilde{U}_{\mu 2} \, \tilde{S}_{21} \,,
    \quad
    S_{\mu\mu} = 
    |\tilde{U}_{\mu 2}|^2 \tilde{S}_{22} + \sum_{\ell\ge 4}
    |\tilde{U}_{\mu\ell}|^2 e^{i\Delta_{\ell\ell}L} \,.
\end{equation}
The expressions for the probabilities, $P_{\alpha\beta} =
|S_{\alpha\beta}|^2$, are straightforward. Defining
\begin{equation}
    d_\mu \equiv \sum_{\ell \ge 4} |U_{\mu\ell}|^2    
\end{equation}
we note that $|\tilde{U}_{\mu 2}|^2 + \sum_{\ell \ge 4}
|\tilde{U}_{\mu\ell}|^2 = 1$ and that $U_{\alpha\ell} =
\tilde{U}_{\alpha\ell}$ for $\ell \ge 4$, so that $|\tilde{U}_{\mu
2}|^2 = 1 - d_\mu$. Therefore:
\begin{gather} \label{eq:Pthree}
    \begin{aligned}
	P_{ee} &= P^{(3)}_{ee} \,,
	\qquad &
	P_{e\mu} &= (1 - d_\mu) \, P^{(3)}_{e\mu} \,,
	\\[1mm]
	P_{\mu e} &= (1 - d_\mu) \, P^{(3)}_{\mu e} \,,
	\qquad &
	P_{\mu\mu} &= (1 - d_\mu)^2 \, P^{(3)}_{\mu\mu}
	+ \sum_{\ell\ge 4} |U_{\mu\ell}|^4 \,,
    \end{aligned}
\end{gather}
where we have used the fact that the terms containing a factor
$e^{i\Delta_{\ell\ell}L}$ oscillate very fast, and therefore vanish
once the finite energy resolution of the detector is taken into
account. In the above expression $P^{(3)}_{\alpha\beta}$ is the
effective probability derived from the Hamiltonian $H^{(3)}$, which
has an ordinary three-neutrino term $H_\text{SM}$ (including the usual
charged-current interaction term of the electron neutrino) plus a
matter term arising from the sterile part of $\tilde{U}^\dagger V
\tilde{U}$:
\begin{equation} \label{eq:Hthree}
    H^{(3)} = H_\text{SM}
    \pm \sqrt{2} \, G_F \, \frac{N_n}{2} \sum_{\alpha=\text{sterile}}
    \begin{pmatrix}
	0 & 0 & 0
	\\
	0 & |\tilde{U}_{\alpha 2}|^2
	& \tilde{U}_{\alpha 2}^* \tilde{U}_{\alpha 3}
	\\
	0 & \tilde{U}_{\alpha 2} \tilde{U}_{\alpha 3}^*
	& |\tilde{U}_{\alpha 3}|^2
    \end{pmatrix} \,.
\end{equation}
Eqs.~\eqref{eq:Pthree} and~\eqref{eq:Hthree} are valid for any number of
extra sterile states.


\section{Robustness of the ATM+LBL bound on $d_\mu$}
\label{app:robust}

As discussed in Refs.~\cite{Bilenky:1999ny, cornering}, the
contribution of atmospheric neutrino data to the disappearance data
set plays a crucial role in rejecting sterile neutrino models. In this
appendix we re-consider the bound on $d_\mu$ in (3+1) schemes,
generalise it to the (3+$s$) case, and investigate the impact of some
of the adopted approximations.

\subsection{Decoupling electron neutrinos}
\label{app:decouple}

\begin{figure}[t] \centering 
    \includegraphics[width=0.85\textwidth]{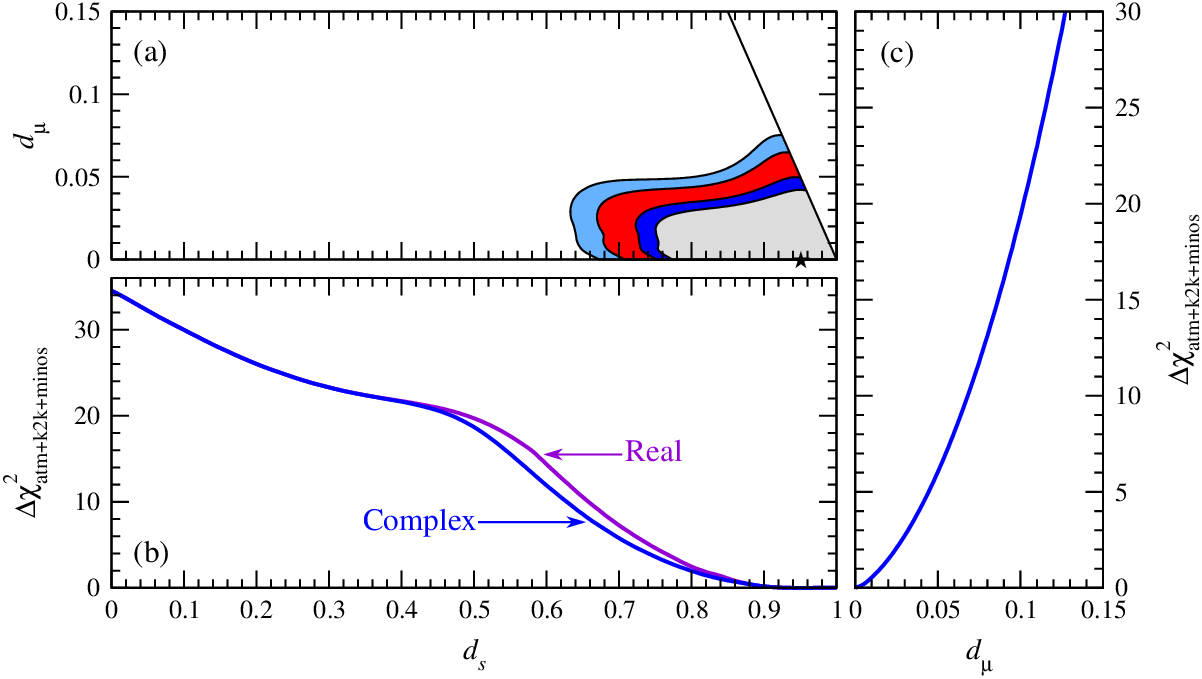}
    \mycaption{\label{fig:atm-4nu}%
    Allowed regions for atmospheric and long-baseline data in (3+1)
    schemes at 90\%, 95\%, 99\%, 99.73\%~CL (2~dof), and projections
    of $\Delta\chi^2$ over $d_s = |U_{s4}|^2$ and $d_\mu = |U_{\mu
    4}|^2$. In panel (b) the purple line (``Real'') is obtained by
    forcing the complex phase $\varphi_{34}$ to $0^\circ$ or
    $180^\circ$, while for the blue line (``Complex'') $\varphi_{34}$
    is left free. In panel (c) the purple line is hidden by the blue
    one.}
\end{figure}

Let us begin by considering the simplified case $\Delta m_{21}^2 = 0$
and $\theta_{13} = 0$. In this limit, the electron neutrino state is
completely decoupled, so that $P_{ee} = 1$ and $P_{e\mu} = P_{\mu e} =
0$.\footnote{Up to now this approximation has been always adopted in
the literature, and in the following Sec.~\ref{app:atm-nue} we are
going to relax it for the first time.} Eqs.~\eqref{eq:Pthree}
and~\eqref{eq:Hthree} then reduce to an effective two-neutrino form in
the $\mu - \tau$ sector:
\begin{gather}
    \label{eq:Ptwo}
    P_{\mu\mu} = (1 - d_\mu)^2 \, P^{(2)}_{\mu\mu}
    + \sum_{\ell\ge 4} |U_{\mu\ell}|^4 \,,
    \\
    \label{eq:Htwo}
    H^{(2)} = \frac{\Dmq_{31}}{4E}
    \begin{pmatrix}
	-\cos2\theta_{23} & \sin2\theta_{23} \\
	\hphantom{-} \sin2\theta_{23} & \cos2\theta_{23}
    \end{pmatrix}
    \pm \sqrt{2} \, G_F \, \frac{N_n}{2} \sum_{\alpha=\text{sterile}}
    \begin{pmatrix}
	|\tilde{U}_{\alpha 2}|^2
	& \tilde{U}_{\alpha 2}^* \tilde{U}_{\alpha 3}
	\\
	\tilde{U}_{\alpha 2} \tilde{U}_{\alpha 3}^*
	& |\tilde{U}_{\alpha 3}|^2
    \end{pmatrix} \,.
\end{gather}

In the case of only one extra sterile state it is possible to perform
a full numerical analysis. The details of such an analysis have been
widely discussed in Refs.~\cite{sol-atm-4nu, MSV-4nu, Maltoni:2004ei},
and are summarised in Fig.~\ref{fig:atm-4nu}. As can be seen from this
figure, atmospheric and long-baseline neutrino data strongly prefer a
pure two-neutrino $\nu_\mu \to \nu_\tau$ oscillation scenario,
disfavouring both a sterile neutrino contribution to the main
$\Dmq_\text{atm}$ oscillations (parametrised by $d_s = |U_{s4}|^2$)
and a mixing of $\nu_\mu$ with the heavy mass eigenstate (parametrised
by $d_\mu = |U_{\mu 4}|^2$).
In this work we are mainly interested in the bound on $d_\mu$, since
the other parameters are ``private'' to atmospheric and long-baseline
data and can therefore be marginalised.

Let us now turn our attention to five-neutrino models. Even in this
case it is possible to perform a full numerical analysis, presented in
Fig.~\ref{fig:atm-dmu}(a). As mentioned in appendix~\ref{app:atm}, in
principle atmospheric and long-baseline data constrain separately
$|U_{\mu 4}|^2$ and $|U_{\mu 5}|^2$, which we parametrise in terms of
$d_\mu$ and $\xi_\mu \equiv (|U_{\mu 4}|^2 - |U_{\mu 5}|^2) / d_\mu$.
However, as can be seen from Fig.~\ref{fig:atm-dmu}(a) the allowed
region is practically independent of $\xi_\mu$. Furthermore, comparing
this figure with Fig.~\ref{fig:atm-4nu}(c) it turns out that the bound
on $d_\mu$ in four-neutrino and five-neutrino models is practically
the same. In other words, the extra freedom which we have in $5\nu$
schemes with respect to $4\nu$ ones is essentially irrelevant for the
constraint on $d_\mu$.

\begin{figure}[t] \centering 
    \includegraphics[width=0.95\textwidth]{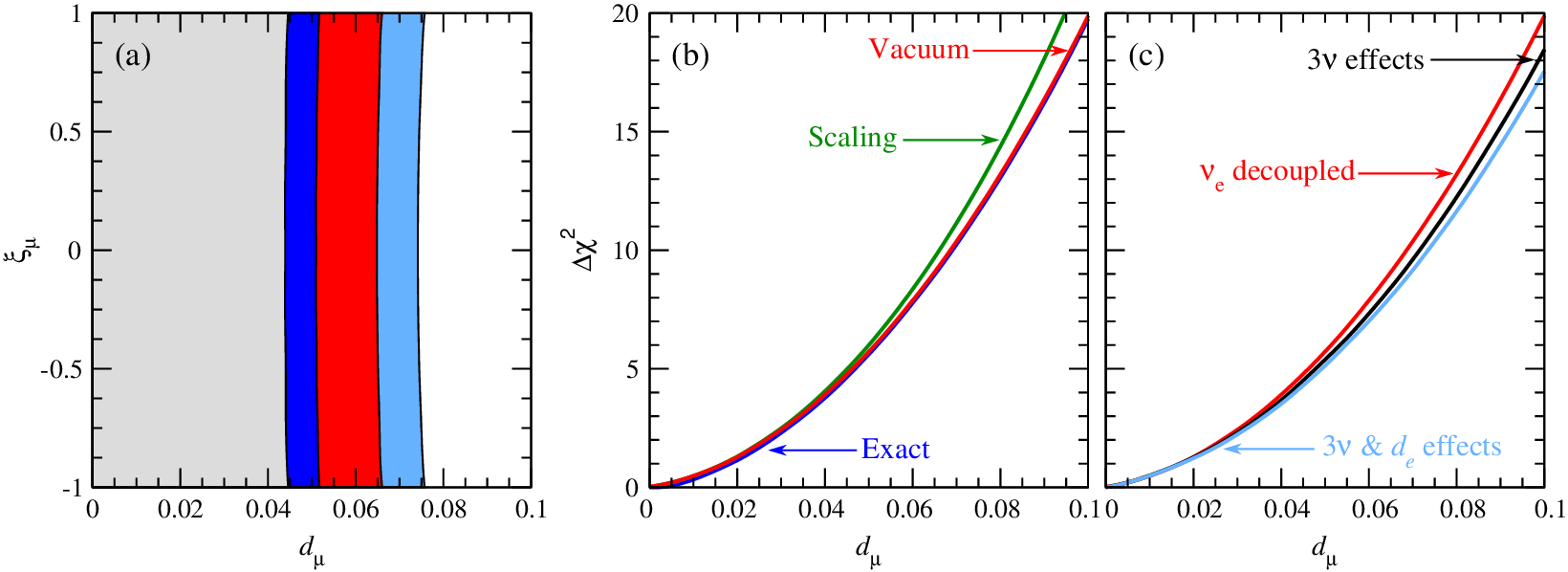}
    \mycaption{\label{fig:atm-dmu}%
      (a) Allowed regions at 90\%, 95\%, 99\%, 99.73\%~CL (2~dof) in
      the $(d_\mu,\, \xi_\mu)$ plane from the analysis of ATM+LBL data
      in (3+2) schemes. (b) \& (c) Impact of different approximations
      on the atmospheric and long-baseline bound on $d_\mu$, see text
      for details.}
\end{figure}

In order to understand this result, let us go back to
Eqs.~\eqref{eq:Ptwo} and~\eqref{eq:Htwo} and consider the differences
between $4\nu$ and $5\nu$ models. They arise from two facts:
\begin{itemize}
  \item In $4\nu$ models we have $P_{\mu\mu} = (1 - d_\mu)^2 \,
    P^{(2)}_{\mu\mu} + d_\mu^2$, so that the ``scaling'' term $(1 -
    d_\mu)^2$ and the ``constant'' term $\sum_{\ell\ge 4}
    |U_{\mu\ell}|^4$ in Eq.~\eqref{eq:Ptwo} are related to each other.
    Conversely, in $5\nu$ models $P_{\mu\mu} = (1 - d_\mu)^2 \,
    P^{(2)}_{\mu\mu} + d_\mu^2 (1 + \xi_\mu^2) / 2$, so that the two
    terms are independent.

  \item The expression for $P_{\mu\mu}^{(2)}$ is different in the two
    models, due to the different contributions to the sterile matter
    term in Eq.~\eqref{eq:Htwo}. Again, in five-neutrino models we
    have more freedom.
\end{itemize}
The relevance of these differences is illustrated in
Fig.~\ref{fig:atm-dmu}(b). All the lines correspond to four-neutrino
models. The blue line (``Exact'') is the exact bound on $d_\mu$ from
atmospheric and long-baseline data, and coincides with the one shown
in Fig.~\ref{fig:atm-4nu}(c). The red line (``Vacuum'') is obtained by
neglecting the last term in Eq.~\eqref{eq:Htwo}, \textit{i.e.}\ by
considering only the vacuum part of $H^{(2)}$. Finally, the green line
(``Scaling'') is obtained by further neglecting the constant term in
Eq.~\eqref{eq:Ptwo}, thus setting $P_{\mu\mu} = (1 - d_\mu)^2 \,
P^{(2)}_{\mu\mu}$.
As can be seen, sterile-induced matter terms are completely irrelevant
for what concerns the bound on $d_\mu$, and the constant term in the
expression for $P_{\mu\mu}$ plays only a minor role. The real bound
arises from the scaling term in $P_{\mu\mu}$, which is the same in
four-neutrino and five-neutrino models. This explains why the
differences between the two schemes are so small. Similarly, the weak
dependence on $\xi_\mu$ in $5\nu$ models arises from the constant term
in Eq.~\eqref{eq:Ptwo}; in particular, both Figs.~\ref{fig:atm-dmu}(a)
and~\ref{fig:atm-dmu}(b) show that as this term is decreased the
quality of the fit gets worse.

In summary, although in principle atmospheric and long-baseline data
could constrain separately all the $|U_{\mu\ell}|^2$ terms in models
with extra sterile states, in practice they are only sensitive to the
sum of these terms, $d_\mu = \sum_{\ell\ge 4} |U_{\mu\ell}|^2$.
Furthermore, the bound on $d_\mu$ is essentially independent on the
number of extra sterile states. The validity of this approximation is
crucial for the analysis of (3+3) models, where an exact treatment
would be very hard to do due to the large number of parameters
involved.

\subsection{Including electron neutrinos}
\label{app:atm-nue}

In the previous section we have seen that the bound on $d_\mu$
reflects the ability of atmospheric and long-baseline data to
effectively fix the total normalisation of $\mu$-like events. This is
possible in spite of the large normalisation errors (of the order of
20\%) on the atmospheric neutrino fluxes and the neutrino-nucleon
cross sections, since the accurate measurements of $e$-like neutrino
events provided by atmospheric data allows to effectively resolve
these uncertainties. In other words, what really matters to determine
the bound on $d_\mu$ is the relative normalisation of $\mu$-like and
$e$-like neutrino events. This opens up the possibility that
subleading effects modifying the distribution of electron neutrino
data may have a sizable impact on the bound on $d_\mu$. In this
section we investigate in detail this possibility.

In the context of sterile neutrino schemes there are two types of
contributions which affect $e$-like events: (i) ``usual''
three-neutrino effects induced by $\Dmq_{21}$ or by $\theta_{13}$, and
(ii) genuine sterile-$\nu$ effects induced by non-zero
$\theta_{1\ell}$ (with $\ell \ge 4$). The formalism to study the
three-neutrino effects has been developed in appendix~\ref{app:atm}.
Following the results of Sec.~\ref{app:decouple}, we assume that the
sterile-induced matter effects in Eq.~\eqref{eq:Hthree} can be
neglected, in which case the effective Hamiltonian $H^{(3)}$ and the
corresponding probabilities $P_{\alpha\beta}^{(3)}$ reduce to the
usual three-neutrino expressions. Note that again the relevant
probabilities are related to the three-neutrino ones only through the
parameter $d_\mu$, and apart from the ``constant'' term in the
expression for $P_{\mu\mu}$ in Eq.~\eqref{eq:Pthree} (see previous
section) these formulas are completely independent from the number of
extra sterile species. The results of our analysis are summarised in
Fig.~\ref{fig:atm-dmu}(c), where we compare the $d_\mu$ bound for the
case when the electron is decoupled (red line) with the same bound
including also three-neutrino effects due to $\theta_{13}$ and
$\Dmq_{21}$ (black line). Note that the Chooz experiment is also
included in the fit. As can be seen from this figure, an accurate
treatment of subleading three-neutrino effects indeed weakens the
bound on $d_\mu$, however the effect is very small and has no impact
on the conclusions of this work.

Let us now study effects induced by non-zero values of
$\theta_{1\ell}$. Following the derivation of appendix~\ref{app:atm},
it is easy to see that in this case $\tilde{U}_{\mu 1}$ and
$\tilde{U}_{\mu 3}$ no longer vanish, in which case
Eq.~\eqref{eq:Pthree} should be replaced with expressions involving
not only $P_{\alpha\beta}^{(3)}$ but also interference terms between
different entries of $\tilde{S}$. However, it is still true that
$\tilde{U}_{e2} = \tilde{U}_{e3} = 0$, so that:
\begin{equation}
    P_{ee} = (1 - d_e)^2 \, P^{(3)}_{ee}
    + \sum_{\ell\ge 4} |U_{e\ell}|^4 \,,
    \quad\text{with}\quad
    d_e \equiv \sum_{\ell \ge 4} |U_{e\ell}|^2 \,.
\end{equation}
Motivated by this result, we approximate the effects of non-zero
$\theta_{1\ell}$ by introducing an independent scaling factor for
electron events:
\begin{equation}
    N_e(d_e) = (1 - 2 d_e) \, N_e(0) \,.
\end{equation}
If $d_e$ is left free to vary without any constraint, the impact on
the bound on $d_\mu$ is dramatic, since in this case electron events
can no longer fix the $20\%$ flux and cross section normalisation
uncertainties. However, the value of $d_e$ is strongly bounded by
Bugey. We have performed a combined analysis of atmospheric+LBL and
SBL data taking into account that both data samples depend on $d_\mu$
and $d_e$ (using the approximation described above). We find that the
final result is practically the same as in our standard case, where
the dependence of the ATM data on $d_e$ is neglected. To illustrate
this we show in Fig.~\ref{fig:atm-dmu}(c), light-blue line, also the
bound on $d_\mu$ from ATM data by adding to the overall $\chi^2$ a
term $(d_e / 0.012)^2$, which simulates roughly the constraint from
Bugey, neglecting that it actually depends on $\Dmq_{41}$. From this
figure it becomes clear that once $d_e$ is limited by the data from
Bugey its impact on atmospheric+LBL data is very small.

In conclusion, the atmospheric bound on $d_\mu$ is robust under our
approximations. However, as clear from the above discussion it depends
on the assumptions about uncertainties on quantities (like fluxes or
cross sections) affecting the ratio of $e$-like and $\mu$-like
atmospheric neutrino event normalisations.


\end{document}